\documentclass[twoside,11pt]{article}

\usepackage[accepted]{melba}
\usepackage{verbatim}
\usepackage{array, multirow, caption, booktabs}
\usepackage{subcaption}
\usepackage[dvipsnames]{xcolor}

\usepackage{mwe} 

\usepackage{amsmath,amsfonts}

\melbaheading{2022:030}{https://www.melba-journal.org/papers/2022:030.html}{2022}{1-33}{05/2022}{12/2022}{Bhat, Pluim, Viergever, and Kuijf}{}{}

\ShortHeadings{Influence of uncertainty estimation techniques on false-positive reduction}{Bhat et al.}
\firstpageno{1}

\title{Influence of uncertainty estimation techniques on false-positive reduction in liver lesion detection}

\author{\name Ishaan Bhat \email i.r.bhat@umcutrecht.nl \\  
	\addr Image Sciences Institute, University Medical Center Utrecht, Heidelberglaan 100, 3584 CX Utrecht, The Netherlands
	\AND
	\name Josien P.W. Pluim \email j.pluim@tue.nl \\
	\addr Department of Biomedical Engineering, Eindhoven University of Technology, Groene Loper 3, 5612 AE Eindhoven, The Netherlands
	\AND
	\name Max A. Viergever \email m.a.viergever@umcutrecht.nl \\
	\addr Image Sciences Institute, University Medical Center Utrecht, Heidelberglaan 100, 3584 CX Utrecht, The Netherlands
	\AND
	\name Hugo J. Kuijf \email h.kuijf@umcutrecht.nl \\
	\addr Image Sciences Institute, University Medical Center Utrecht, Heidelberglaan 100, 3584 CX Utrecht, The Netherlands
}

\begin{document}

\maketitle

\begin{abstract}
Deep learning techniques show success in detecting objects in medical images, but still suffer from false-positive predictions that may hinder accurate diagnosis. The estimated uncertainty of the neural network output has been used to flag incorrect predictions. We study the role played by features computed from neural network uncertainty estimates and shape-based features computed from binary predictions in reducing false positives in liver lesion detection by developing a classification-based post-processing step for different uncertainty estimation methods. We demonstrate an improvement in the lesion detection performance of the neural network (with respect to F1-score) for all uncertainty estimation methods on two datasets, comprising abdominal MR and CT images, respectively. We show that features computed from neural network uncertainty estimates tend not to contribute much toward reducing false positives. Our results show that factors like class imbalance (true over false positive ratio) and shape-based features extracted from uncertainty maps play an important role in distinguishing false positive from true positive predictions. Our code can be found at \url{https://github.com/ishaanb92/FPCPipeline}
\end{abstract}


\section{Introduction}
\label{sec:introduction}

Deep learning systems have become the go-to approach for various medical image analysis tasks~\citep{zhou2021_review, litjens_survey_2017}. However, such systems may make erroneous predictions. These arise due to a variety of reasons, for example, the model overfitting to the training data, presence of noise/artefacts in the image, or mismatch between the training and clinical data distributions~\citep{castro_causality_2020}. False-positive prediction is one such type of error and may hinder accurate diagnosis. 

Lesion detection in medical imaging is usually formulated as an image segmentation problem, where labels are assigned to each voxel in the image. The task suffers from class imbalance, because lesions occupy a small volume in the image. The imbalance is usually corrected by re-weighting components of the loss function and/or selectively sampling data. While this may lead to more voxels being correctly classified as lesions, the prediction algorithm may become biased toward detecting more lesions than are actually present. False-positive reduction is accomplished via negative sampling i.e. sampling background to reduce algorithmic bias toward finding lesions~\citep{ciga_overcoming_2021}, architectural modifications~\citep{tao_new_2019}, or by using a classifier as a post-processing step~\citep{chlebus_automatic_2018, bhat_using_2021}.  

In this paper, we extend initial work performed in \cite{bhat_using_2021} by developing a system to reduce false-positive predictions using uncertainty estimations made by the neural network. We study the effect uncertainty estimation methods have on the detection of false positive lesions. To do so, we develop a model-agnostic post-processing pipeline that can work with predictions made by any probabilistic classifier.

We summarize related work in Section \ref{sec:rel_work}.  Section \ref{sec:data} describes the datasets used in this paper. In Section \ref{sec:mm} we provide background about the different uncertainty methods we studied, and  describe the false-positive classification pipeline. In Section \ref{sec:exp} we describe the neural network architecture and training (Section \ref{sec:nn_training}), and the different evaluation methods used to analyse the performance of uncertainty estimation methods to reduce false positives (Section \ref{sec:eval}). Results are shown in Section \ref{sec:results} and their implications are discussed in Section \ref{sec:discussion}.

\section{Related Works}
\label{sec:rel_work}
The estimation of prediction uncertainty has been useful in domains such as detecting adversarial~\citep{smith_understanding_2018} and out-of-distribution~\citep{ovadia_can_2019} examples, robotics~\citep{loquercio_general_2020}, autonomous driving~\citep{hoel2020}, and reinforcement learning~\citep{depeweg_decomposition_2018}. Uncertainty estimation has also found numerous promising applications in the field of medical image analysis~\citep{ching_opportunities_2018}.

Uncertainty estimates have been used as a proxy to determining the quality of the predictions, with poor quality predictions being referred to an expert. \cite{leibig_leveraging_2017} showed that the detection of diabetic retinopathy improved when predictions with estimated uncertainty above a certain threshold were referred to an expert. \cite{seebock_exploiting_2020} used the uncertainty map to detect anomalous regions in retinal OCT images. For the task of cardiac MR segmentation, \cite{sander_towards_2019} referred voxels with high uncertainty to an expert to produce segmentations with more accurate boundaries. \cite{stoyanov_joint_2018} showed the benefits of using the uncertainty map as a visual aid to clinicians for the task of retinal layer segmentation in OCT images. \cite{camarasa2021} show metrics derived from the uncertainty maps can be used to evaluate the quality of multi-class segmentations of the carotid artery. \cite{karimi_accurate_2019} used uncertainty estimation to flag poor segmentations and used a statistical shape model to improve the result for the task of prostate segmentation from ultrasound images. In addition to determining quality of predictions, uncertainty estimates have also been shown to improve performance of downstream tasks within a cascade~\citep{greenspan_propagating_2019}. 

When it comes to segmenting structures of interest within images, such as lesions, there is a need to quantify uncertainty at the structure level, by aggregating per-voxel uncertainty estimates, because it is at the structure level that diagnosis is performed. \cite{nair_exploring_2020} showed that the log-sum of voxel-wise uncertainties computed over the detected structures can be used to filter small false positives. Similarly, \cite{graham_mild-net_2019} used the mean uncertainty over segmented glandular structures to disregard poor quality predictions made on colon histopathological images. \cite{mehrtash_confidence_2019} showed that for prostate, brain lesion and cardiac segmentation, the mean uncertainty over the segmented structure had a negative correlation with the Dice overlap and therefore the mean uncertainty could be used as a proxy for segmentation quality. Similarly, \cite{ng_estimating_2020} and \cite{roy_inherent_2018} developed metrics to check variations in the prediction at the structure level and showed a negative correlation between this metric and the Dice coefficient for cardiac and brain lesion segmentation, respectively. \cite{eaton-rosen_towards_2018} used uncertainty estimation to compute well-calibrated error bars for the estimated brain lesion volumes for clinical use.  

There is also a class of methods that use uncertainty estimates as an input to a second stage that is used to predict segmentation quality or detect segmentation failures. \cite{jungo_analyzing_2020} used the task of segmenting different types of lesions from brain MR images to show that aggregating features computed from the per-voxel uncertainty maps can predict segmentation quality. These features correlated well with the Dice score of the prediction and therefore could be used to a flag a poor segmentation. In a similar vein, \cite{devries_leveraging_2018} used a second neural network as a regressor to predict the Jaccard index using the per-voxel uncertainty map and binary segmentation as inputs for their skin lesion segmentation task. \cite{sander_automatic_2020} used a second neural network to detect local segmentation errors in a cardiac segmentation task using the uncertainty map, the prediction, and image patches as inputs.


Most approaches use voxel-wise uncertainty estimates to detect segmentation failures, or use them as a proxy for segmentation quality. ~\cite{nair_exploring_2020} showed that aggregating voxel-wise uncertainty estimates over the predicted lesion volume using the log-sum operation can be used to filter small false positives. \cite{jungo_analyzing_2020} compared different ways to perform spatial aggregation at the subject level and concluded using radiomics features to construct subject-level feature vectors performed best at detecting failures while segmenting tumours in brain MR images. Since this is a more general form of spatial aggregation, we construct radiomics feature vectors corresponding to predicted lesion volumes from the uncertainty map estimated by the neural network to investigate the role of uncertainty estimates in reducing false-positive detections.

\section{Data}
\label{sec:data}

\subsection{UMC MR dataset}
We used abdominal dynamic contrast enhanced (DCE) and diffusion weighted (DW) MRI of 71 patients with liver metastases from the University Medical Center Utrecht, the Netherlands. This data was acquired between February 2015 and February 2018. The DCE-MR series was acquired in six breath holds, resulting in a total of 16 3-D contrast phases per patient. Voxel spacing for these images is $1.5\,\mathrm{mm} \times 1.5\,\mathrm{mm} \times 2\,\mathrm{mm}$. The liver and the metastases within the liver were manually segmented on the DCE-MRI by a radiologist in training and verified by a radiologist with more than 10 years of experience. The dataset mainly included colorectal and neuroendocrine metastases, with few other types in addition (i.e., other gastrointestinal metastases and breast metastases). Motion correction between the different contrast phases for each DCE MR image was performed using the techniques described in \cite{jansen_evaluation_2017}.

The DW-MR images were acquired with three b-values: $10$, $150$, and $1000$ s/$\mathrm{mm}^2$, using a protocol with the following parameters: TE: $70$ ms; TR: $1.660$ ms; flip angle: $90$ degrees. For each patient, the DW MR image was non-linearly registered to the DCE MR image using the elastix~\citep{elastix2010} toolbox.

We applied the manually created liver masks to the abdominal DCE and DW MR images. All images were resized to in-place dimensions of $256 \times 256$, followed by z-score normalization of the intensities. The 16 contrast phases of the DCE MR images were divided into groups and averaged to generate a 6-channel image~\citep{jansenspie}. The contrast phases were grouped as follows, the pre-contrast image, the early arterial phase, the late arterial phase, the hepatic/portal-venous phase, the late portal-venous/equilibrium phase, and the late equilibrium. The DW images were concatenated with the DCE images along the channel axis to create a 9-channel input image for the neural network. 

The data was split into 50 training patients, 5 validation patients and 16 test patients.

\subsection{Liver lesion Segmentation Challenge (LiTS)}

The LiTS dataset~\citep{bilic_liver_2019}\footnote{\url{https://competitions.codalab.org/competitions/17094}} contains 131 abdominal CT images with reference liver and lesion segmentations from seven hospitals and research institutes. Out of these 131 images, we excluded 13 images that had no lesions. All images in the dataset had in-plane dimensions of $512$ x $512$ with variable voxel spacings. The median voxel spacing for the dataset was $0.76$ x $0.76$ x $1.0$ mm.

Pre-processing on the images was performed by first clipping the intensities to the range $[-100, 200]$ and then re-scaling the clipped intensities to the range $[0,1]$. Non-liver regions in the image were masked using the reference liver segmentation. 

We divided our reduced dataset of 118 CT images into non-overlapping sets with 94 training, 6 validation and 18 test patients. 

\section{Materials and Methods}
\label{sec:mm}

We have developed a classification pipeline to study the effect of uncertainty estimation techniques on false-positive reduction. Section \ref{sec:theory} provides a general framework to describe different uncertainty estimation methods. Section \ref{subsec:lesion_det} explains our graph-based lesion counting method to account for many-to-one and one-to-many correspondences between the predicted and ground truth segmentations while evaluating lesion detection performance. Finally, Section \ref{subsec:fp_class} describes the classification pipeline used to reduce false-positive lesion segmentations.

\subsection{Theory}
\label{sec:theory}
Standard neural network training and inference provide point estimates for their prediction. There has been early work on uncertainty estimation utilizing a Bayesian framework to train neural networks that can produce a distribution instead of a point estimate~\citep{mackay_bayesian, neal_bayesian}. However, neural network training and inference with these approaches was computationally expensive. Recent approaches build upon these earlier approaches to make them computationally efficient using variational inference techniques~\citep{blundell_weight_2015, gal_dropout_2015, kingma_variational_2015}, the most widely adopted amongst these being \emph{MC-Dropout}~\citep{gal_dropout_2015}.
In addition to these methods, techniques such as using model ensembles~\citep{lakshminarayanan_simple_2017} and test-time augmentation~\citep{wang_aleatoric_2019} are widely used and have been described later in this subsection.

We first present a general method to compute the output distribution of a neural network, and then show how each of the methods mentioned before can be derived from this general formulation. In a Bayesian framework, the output distribution of a neural network, $p(y^*| x^*, X, Y)$, is obtained by marginalizing over the posterior distribution of the model parameters and image transformations:
\begin{equation}
\label{eq:marg_likelihood}
    p(y^*| x^*, X, Y) = \int_{\omega \in \Omega} \int_{\tau \in \Lambda} p(y^*|x^*, \omega, \tau) p(\omega | X, Y)p(\tau)d\tau d\omega
\end{equation}
Here $x^*$, $y^*$ are the input and output values respectively, $X$ and $Y$ are the data and labels used for training respectively, and $\omega$, $\tau$ are the set of neural network parameters and input transformations respectively.
The integral in Equation \ref{eq:marg_likelihood} is approximated using a Monte-Carlo simulation:
\begin{equation}
\label{eq:mc_approx}
 p_{\mathrm{approx}}(y^*|x^*, X, Y) = \frac{1}{NM}\sum_{i=1}^{N}\sum_{j=1}^{M}p(y^*|x^*, \omega_{i}, \tau_j)
\end{equation}

\subsubsection{Baseline}

As a baseline, we included a convolutional neural network with a softmax output producing per-voxel point probability estimates, $p(y^* | x^*, \omega^*)$, where $y^*$, $x^*$ are the output and input respectively, and $\omega^*$ is the set of neural network parameters obtained via optimization. Therefore, the distribution over the model parameters, $p(\omega| X, Y)$ becomes a Dirac delta function at $\omega^*$, and the distribution over input transformations, $p(\tau)$ a Dirac delta function at the identity transform.

\subsubsection{MC-Dropout}

\emph{Dropout}~\citep{srivastava_dropout_2014} is a regularization technique where different random subsets of the neural network weights are set to zero during training, based on sampling masks from a Bernoulli distribution with a fixed probability or \emph{dropout rate}. \cite{gal_dropout_2015} show that weights retained during dropout can be used as samples of the posterior distribution over the weights, $p(\omega| X, Y)$.

Therefore, for MC-Dropout, Equation \ref{eq:mc_approx} reduces to:
\begin{equation}
\label{eq:mc_dropout}
 p_{\mathrm{mcd}}(y^*|x^*) = \frac{1}{N_{\mathrm{mcd}}}\sum_{i=1}^{N_{\mathrm{mcd}}} p(y^*|x^*, \omega_{i})
\end{equation}
Here $\omega_i$ are the model parameters retained during the $i^{th}$ pass during inference.

\subsubsection{Model Ensembles}

Model ensembles~\citep{lakshminarayanan_simple_2017} have been shown to have superior performance and yield high quality uncertainty estimates as compared with a single neural network. Models within an ensemble are trained independently of each other, and the final prediction is obtained by averaging the outputs of the different models. From a Bayesian viewpoint, this can be seen as another way to draw samples from $\Omega$. Therefore, for model ensembles, Equation \ref{eq:mc_approx} becomes:
\begin{equation}
\label{eq:ensemble}
 p_{\mathrm{ensemble}}(y^*|x^*, X, Y) = \frac{1}{N_{\mathrm{ensemble}}}\sum_{i=1}^{N_{\mathrm{ensemble}}} p(y^*|x^*, \omega_{i})
\end{equation}
Here $\omega_{i}$ is the set of weights belonging to the $i^{th}$ model in the ensemble.

\subsubsection{Test-time augmentation}

Test-time augmentation (TTA)~\citep{wang_aleatoric_2019} computes the output probability distribution by marginalizing over a distribution of image transformations, $p(\tau)$, used to transform the input image. 
\begin{equation}
\label{eq:tta_likelihood}
    p_{\mathrm{tta}}(y^*|x^*, \omega^*) = \frac{1}{M_{\mathrm{tta}}}\sum_{j=1}^{M_{\mathrm{tta}}}p(y^*|x^*, \omega^*, \tau_j))
\end{equation}
Here $\tau_j$ is the $j^{th}$ transform sampled from $p(\tau)$ and $\omega^*$ is the point estimate of the model parameters obtained via optimization.

\subsubsection{Combining any method with TTA}

MC-Dropout or model ensembles can be combined with test-time augmentation, as shown in Equation \ref{eq:mc_approx} where $\omega_i$ could either be the weights retained during $i^{th}$ dropout pass or weights from the $i^{th}$ model in the ensemble

\subsubsection{Uncertainty estimation and decoupling}

The predictive uncertainty for classification problems, i.e. with a discrete set of $C$ labels as possible output values, is quantified by the entropy of the output distribution. The entropy for such a distribution $p$ is given by:
\begin{equation}
\label{eq:pred_unc}
u_{\mathrm{predictive}} = H[p] = -\sum_{c=1}^{C}p(y^* = c | x^*)\mathrm{log}(p(y^* = c | x^*))
\end{equation}

Entropy is low when all the probability mass is concentrated on a single class, and high when it is distributed more evenly over classes.

The predictive uncertainty can be decomposed into uncertainty originating from the model (epistemic) and uncertainty originating from the data (aleatoric)~\citep{kiureghian_aleatory_2009}. This decomposition is possible if there is variation in the model parameters used to generate the prediction, as is the case with MC-Dropout or model ensembles~\citep{smith_understanding_2018}.

Aleatoric uncertainty reflects the true noise in the data and cannot be reduced by using more data to train the model. Inherent noise in the data may be reflected in a prediction with high entropy. Thus, the aleatoric uncertainty is the mean entropy computed over predictions with fixed parameters:
\begin{equation}
\label{eq:al_unc}
    u_{\mathrm{aleatoric}} = \frac{1}{T}\sum_{i=1}^{T}H[p(y^*|x^*, \omega_i)]
\end{equation}

The epistemic uncertainty in the prediction is the mutual information between the model parameters and the data~\citep{smith_understanding_2018}. This is equivalent to the epistemic uncertainty being the difference between the predictive uncertainty (Equation \ref{eq:pred_unc}) and the aleatoric uncertainty (Equation \ref{eq:al_unc}):
\begin{equation}
\label{eq:epi_unc}
u_{\mathrm{epistemic}} = u_{\mathrm{predictive}} - u_{\mathrm{aleatoric}}
\end{equation}

For models that use a single fixed set of parameters (eg: TTA), Equation \ref{eq:al_unc} shows that the aleatoric uncertainty is equal to the predictive uncertainty, i.e. the predictive uncertainty is estimated from the data only. For predictions from either MC-Dropout or model ensembles, all three uncertainty measures may be computed.

\begin{figure}[htb]
    \centering
    \includegraphics[width=0.6\linewidth]{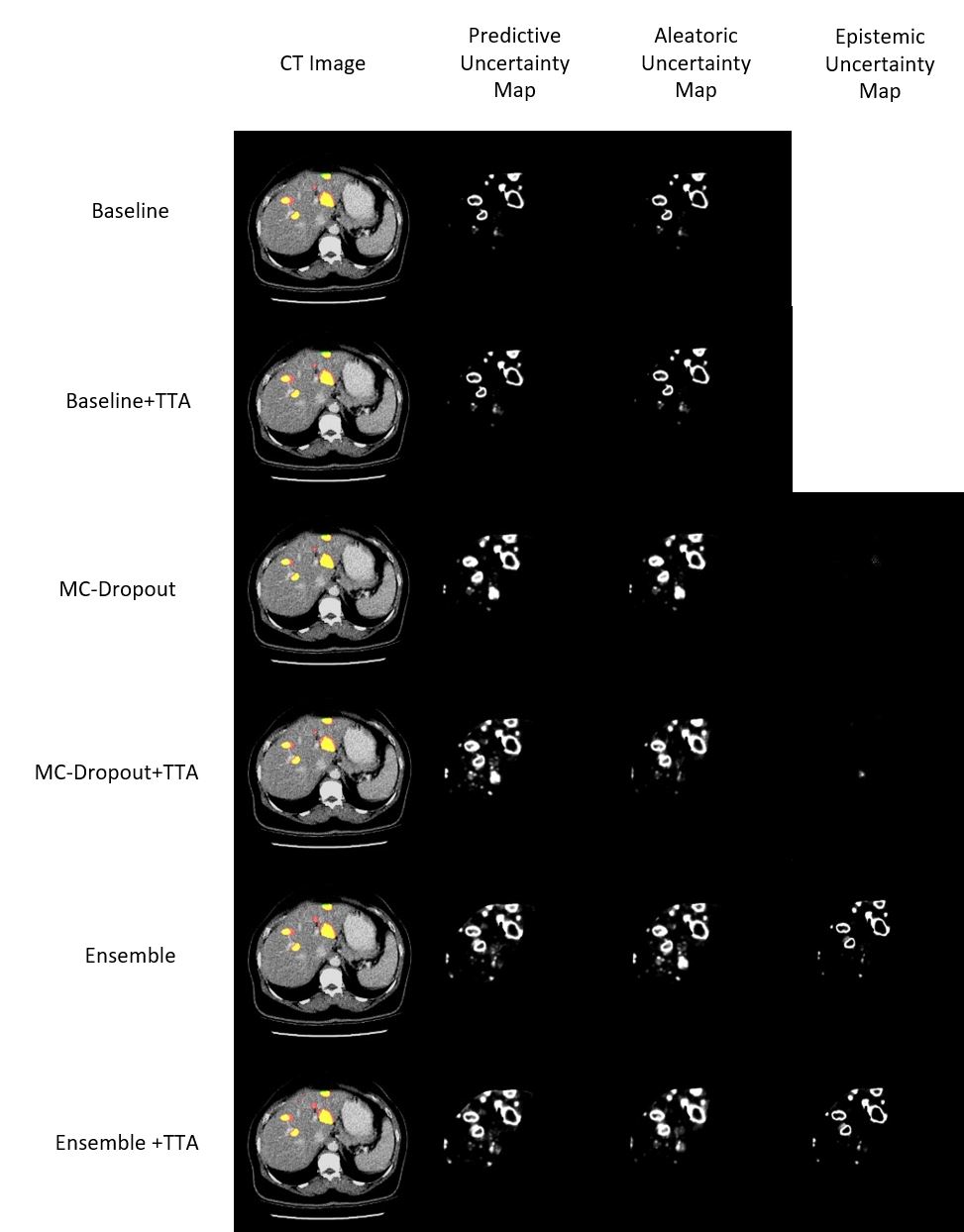}
    \caption{Uncertainty maps for different estimation techniques. In the left-most column, we show the CT image slice with the reference segmentation (green), predicted segmentation (red) and their overlap (yellow). Since model parameters are fixed for Baseline and Baseline+TTA configurations, the epistemic uncertainty is always 0.}
    \label{fig:unc_types}
\end{figure}

Image segmentation can be viewed as a per-voxel classification problem, and per-voxel uncertainty estimates form an uncertainty map. As an example, we show the original image (with the binary prediction overlaid) and uncertainty maps for a single CT image slice from the LiTS dataset in Figure \ref{fig:unc_types}. The epistemic uncertainty maps for MC-Dropout and MC-Dropout+TTA are almost completely empty. This signifies that there is low variation in the MC-Dropout and MC-Dropout+TTA outputs, produced by different passes through the model for a given input (See Equations 6, 7, 8). On the other hand, in Figure \ref{fig:unc_types}, the rows corresponding to ensemble and ensemble+TTA, the epistemic uncertainty maps show signal outlining the lesion boundaries, comparable to the predictive and aleatoric uncertainty maps. Following a similar reasoning, this indicates that there is high variation in the outputs of the different models in the ensemble for a given input. This occurs because in an ensemble of independently trained models, the outputs from each model are less correlated as compared to MC-Dropout outputs, which are generated by sampling subsets of weights optimized jointly~\citep{fort_deep_2020}.

\subsection{Lesion detection}
\label{subsec:lesion_det}
Lesions were detected using a neural network trained to perform image segmentation (See Section \ref{sec:nn_training} for details). Inference was performed on 2-D slices of the images, which were combined to form the segmented volume. For MC-Dropout and test-time augmentation, we used values of $N_{mcd} = 10$ and $M_{tta} = 10$. Larger values of $N_{mcd}$ and $M_{tta}$ did not result in any change in the precision and recall metrics on the test set (shown in Table \ref{tab:lesion_det_met}). For test-time augmentation, random rotations between $[-45, 45]$ degrees were applied to compute predictions. For the ensemble configurations, we used five independently trained models ($N_{ensemble} = 5$) to compute the prediction on test images. We applied a threshold of $0.5$ to convert the per-voxel probability output maps to binary lesion masks. 

Post-processing was performed on the binary lesion masks using a morphological closing with a $3 \times 3 \times 3$  cube-shaped element  followed by morphological opening with a $3 \times 3 \times 3$ cross-shaped element~\citep{jansenspie}. Within the post-processed volume, we performed connected component analysis to identify lesion volumes. 

To quantify the efficacy of our segmentation network at lesion detection, we used the precision, recall and F1 metrics. However, to compute these metrics, it is necessary to find correspondences between the detected lesion volumes and the lesion volumes present in the reference segmentation. Situations could arise where many-to-one or one-to-many correspondences between lesion volumes are found, and any method used to count the number of detected lesions must take this into account~\citep{crimi_dice_2018}.

We tackled the problem of counting detections by constructing a directed bipartite graph with 2 sets of nodes, i.e. the lesions in the predicted segmentation and reference segmentation respectively. We constructed edges between the lesions from the two sets of nodes and set the edge weight between them to be the Dice overlap between the lesion volumes. Edge construction was done in both directions, overlap for a single volume in the predicted segmentation was computed with each object in the reference segmentation and vice versa.

\begin{figure*}[htb]
\centering
\begin{subfigure}[t]{0.45\linewidth}
\includegraphics[width=0.75\linewidth]{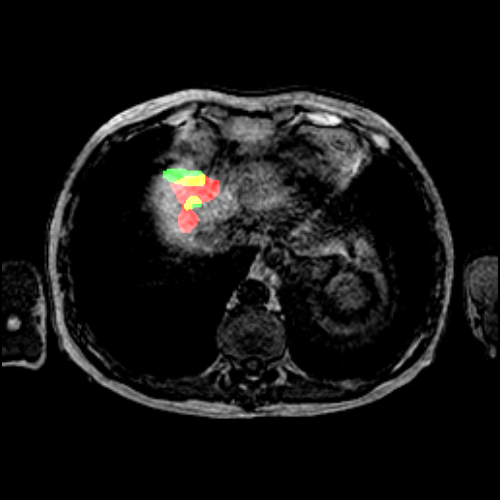}
  \caption{Prediction (red) and reference (green) segmentations (overlap in yellow), overlaid on a DCE MR image. In this image slice, it can be seen that the same predicted lesion volume overlaps with two lesions from the reference segmentations. In such a case, we count two detected lesions.}
  \label{fig:seg_overlay}
\end{subfigure}
 \hspace{1em}
\begin{subfigure}[t]{0.45\linewidth}
  \includegraphics[width=0.75\linewidth]{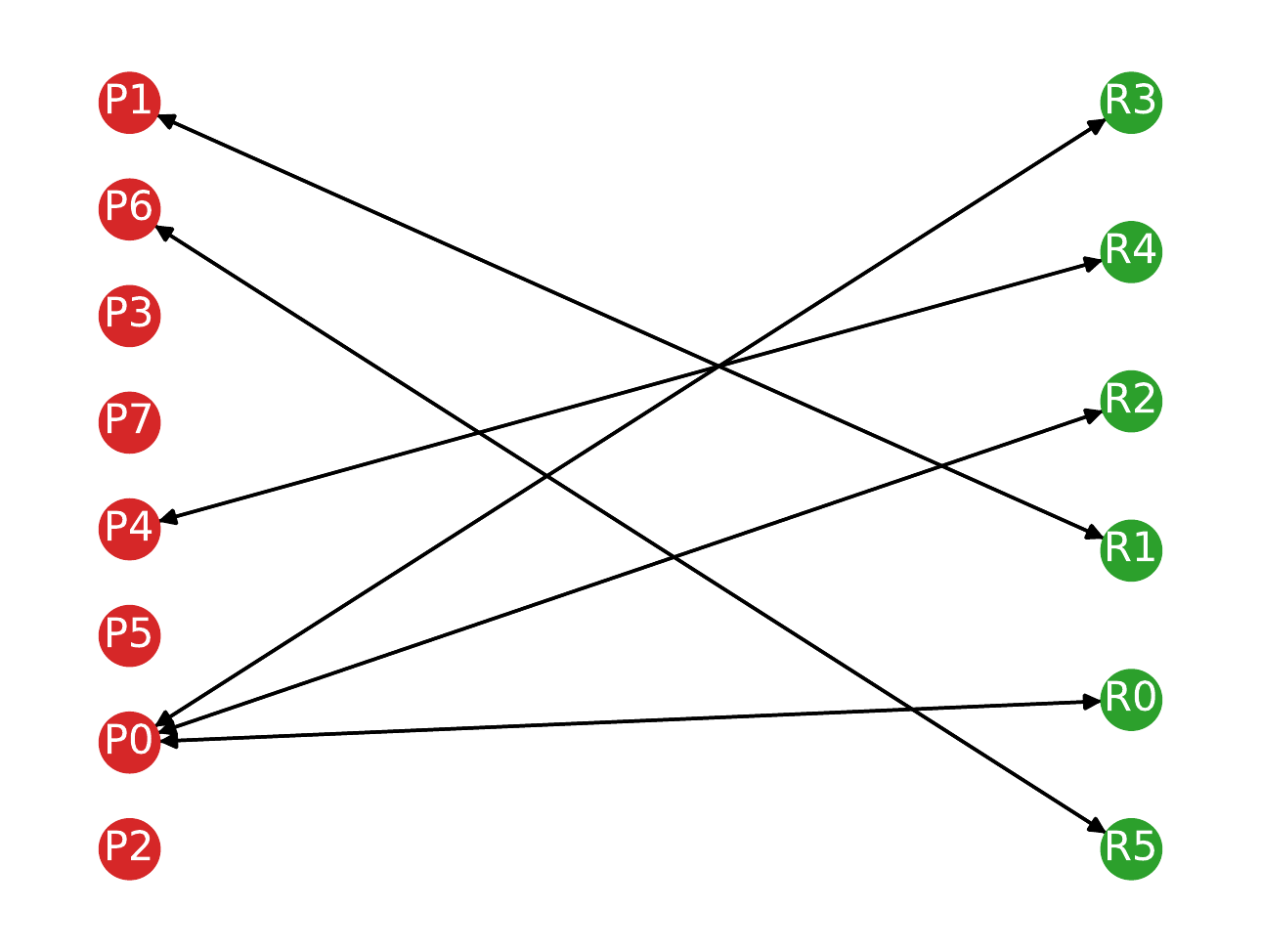}
  \caption{Graph capturing lesion correspondences for the DCE MR image in Figure \ref{fig:seg_overlay} in terms of overlap. The nodes on the left in red are lesions in the predicted segmentation, and those on the right in green are lesions from the reference segmentation. Edges with 0 weight have not been shown for clarity. It can be seen that three lesions (two of which can be seen in Figure \ref{fig:seg_overlay}) in the reference segmentation correspond to a single lesion (P0) in the prediction.}
  \label{fig:lesion_corr_graph}
\end{subfigure}
\caption{Correspondences between predicted and reference segmentations}
\label{fig:lesion_det}
\end{figure*}

Once the graph edges are constructed, counting the number of true positive, false positive and false negative predictions is straightforward. Any volume in the predicted segmentation is considered a \emph{true positive} detection if it has at least one outgoing edge with a non-zero weight. To account for many-to-one correspondences, we count true positives as the number of nodes in the reference lesion partition with at least one incoming edge to avoid multiple counting. A volume in the predicted segmentation is considered a \emph{false positive} prediction if it has no edges with a non-zero weight. Similarly, a volume in the reference segmentation is considered a \emph{false negative} if it has no incoming edges with a non-zero weight. In Figure \ref{fig:lesion_det}, we show an example of a correspondence graph with a many-to-one lesion correspondence.

The detection metrics are computed as follows:

\begin{equation}
    \mathrm{precision} = \frac{\mathrm{true\, positives}}{\mathrm{true\, positives} + \mathrm{false\, positives}}    
\end{equation}

\begin{equation}
\mathrm{recall} = \frac{\mathrm{true\, positives}}{\mathrm{true\, positives} + \mathrm{false\, negatives}}    
\end{equation}

\begin{equation}
    \mathrm{F1} = \frac{2\times\mathrm{precision}\times\mathrm{recall}}{\mathrm{precision} + \mathrm{recall}}
\end{equation}

\subsection{False-positive classification}
\label{subsec:fp_class}
\begin{figure}[htb]
    \centering
    \includegraphics[width=\linewidth]{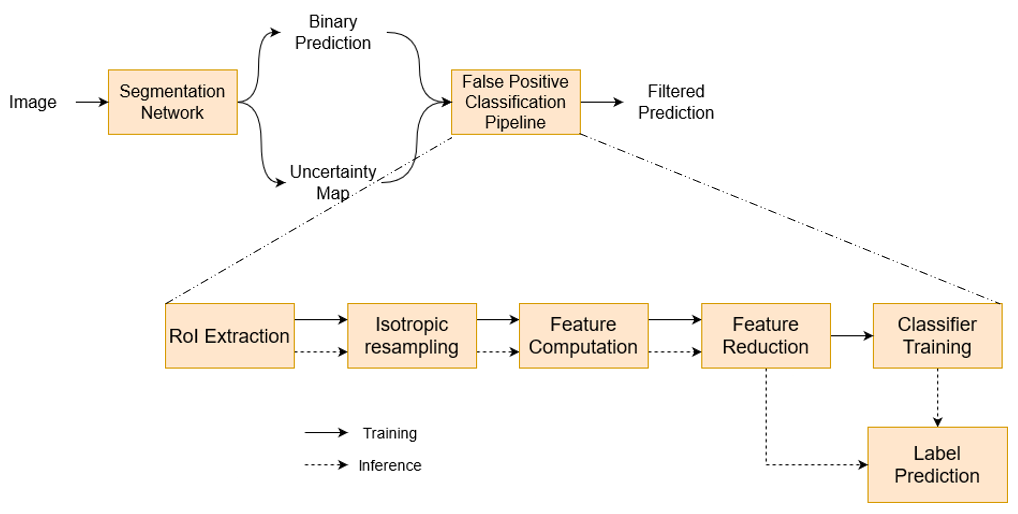}
    \caption{Pipeline to filter false positive lesion predictions from neural network predictions using uncertainty maps.}
    \label{fig:meth}
\end{figure}

The steps involved in filtering false positive predictions from the neural network prediction are shown in Figure \ref{fig:meth}. The pipeline was designed following guidelines on radiomics workflows detailed in \cite{van_timmeren_radiomics_2020}.

We used the lesion detections from the neural network training and validation patients as training data for our false-positive classifier and use the held-out test set to report the performance. 

\subsubsection{Region of interest (RoI) extraction}

We use the binarized neural network prediction to extract 3-D patches from the uncertainty map corresponding to the detected lesions. The lesion correspondence graph is used to assign a label (true positive or false positive) to each patch. The 3-D patch from the uncertainty map(s) and the RoI extracted from the binarized prediction are used for computing intensity and shape features, respectively.

\subsubsection{Isotropic resampling}

We resample each 3-D uncertainty patch and RoI mask to isotropic spacing using $3^{rd}$ order B-spline and nearest-neighbor interpolation, respectively. Isotropic resampling was performed to ensure rotational invariance of texture features~\citep{zwanenburg_image_2020}. For the LiTS dataset, patches were resampled along all axes to $0.76 \mathrm{mm} \times 0.76 \mathrm{mm} \times 0.76 \mathrm{mm}$, the median in-plane voxel spacing for the dataset. For the UMC dataset, we keep the in-plane voxel spacing unchanged and resample in the axial direction to achieve an isotropic spacing of $1.543 \mathrm{mm} \times 1.543 \mathrm{mm} \times 1.543 \mathrm{mm}$.

\subsubsection{Feature Computation}
For each 3-D patch from the uncertainty map, we computed features as described by the Imaging Biomarker Standardization Initiative (IBSI)~\citep{zwanenburg_image_2020}. We used the PyRadiomics~\citep{van_griethuysen_computational_2017} library to compute 99 radiomics features for each patch.

\subsubsection{Feature Reduction}
\begin{figure*}[htb]
\centering
\includegraphics[width=1.0\linewidth]{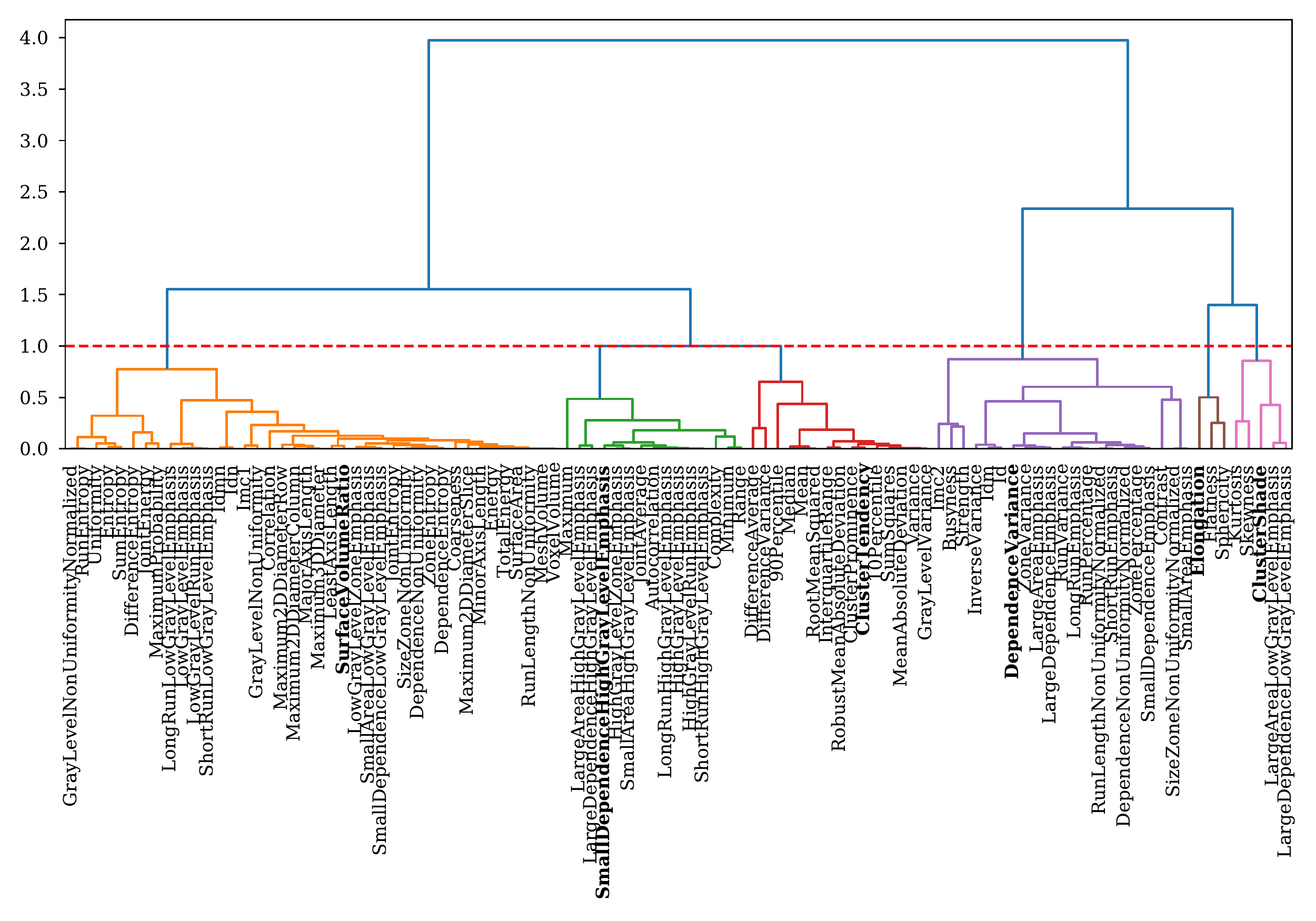}
  \caption{Hierarchical cluster formation based on the feature correlation matrix. The dashed red line shows the threshold at which the dendogram is \emph{cut}. A single feature is selected (based on maximum mutual information with respect to the label) from each of the flat clusters  formed after thresholding (shown in bold).}
  \label{fig:full_features}
\end{figure*}

We designed our feature reduction step to produce a set of minimally correlated features. Hierarchical clustering was performed by constructing a distance matrix such that distance between a pair of features was inversely proportional to their correlation. In Figure \ref{fig:full_features} we show the hierarchical clusters formed for the MC-Dropout (Predictive) configuration of the LiTS dataset. The next step was forming \emph{flat} clusters of features by choosing a threshold at which the dendogram is \emph{cut}. All leaves below a cut form a single \emph{flat} cluster, and the feature with maximum mutual information with respect to the labels was chosen from each of the flat clusters (highlighted in bold in Figure \ref{fig:full_features}). The same threshold ($1.0$) was used for both datasets and all configurations to produce an almost uniform number of features. The dendogram threshold was chosen by visually examining hierarchical clusters for all configurations and datasets. The threshold was chosen to ensure approximately the same number and sizes of clusters. This could be achieved for most configurations, although some were outliers with fewer features. The idea of using a fixed threshold was to ensure the use of the same extent of correlation between features while converting the hierarchical clusters to flat clusters.

In Table \ref{tab:opt_feat} we show reduced feature vector sizes ($N^*$) computed for all the configurations for both datasets.

\begin{table}[htb]
\centering
\caption{Reduced feature vector sizes across datasets and configurations}
\resizebox{0.6\textwidth}{!}{\begin{tabular}{cccc}
\toprule
{} & {} & {LiTS} & UMC\\
 \cmidrule(lr){3-3}
 \cmidrule(lr){4-4}
 Configuration & Uncertainty Type & N\textsuperscript{*} & N\textsuperscript{*} \\
   \midrule
Baseline & Predictive & 9 & 9 \\
Baseline + TTA & Predictive & 9 & 8 \\
\midrule
 & Predictive & 6 & 7 \\
MC-Dropout & Aleatoric & 5 & 7 \\
 & Epistemic & 6 & 6 \\
\midrule
 & Predictive & 5 & 7 \\
MC-Dropout + TTA & Aleatoric & 5 & 7 \\
 & Epistemic & 6 & 5 \\
\midrule
 & Predictive & 8 & 8 \\
Ensemble & Aleatoric & 9 & 9 \\
 & Epistemic & 7 & 8 \\
\midrule
 & Predictive & 8 & 8 \\
Ensemble + TTA & Aleatoric & 9 & 8 \\
 & Epistemic & 8 & 8 \\
 \bottomrule
\end{tabular}}
\label{tab:opt_feat}
\end{table}

\subsubsection{Classifier Training}
We used features computed from uncertainty map patches, corresponding to lesion detections in the training and validation images for the neural network, to train an Extremely Randomized Trees (ERT)~\citep{geurts2006} classifier to classify the detected lesion, as a true or false positive. This classifier is an example of a bagging predictor~\citep{breiman_bagging_1996}, which performs the classification task using an ensemble of randomized base classifiers. The base classifier for the ERT is a decision tree, and randomization is introduced in each decision tree using two methods. First, random subsets of input features are used to split tree nodes. Second, thresholds for each of the input features are sampled randomly and the best of these is used as the splitting rule. Injecting randomness in the base classifiers decouple their prediction errors and reduce overfitting in the constructed ensemble.

The hyperparameters for the classifier were chosen via five-fold cross-validation while optimizing for the area under the curve (AUC) metric, to achieve a good trade-off between sensitivity and specificity. The hyperparameters are shown in Table \ref{tab:class_hyperparams} in Appendix \ref{app:class_hyperparams}. The number of training samples for both datasets and all uncertainty estimation methods is shown in Table \ref{tab:samples_tab}. We used the scikit-learn~\citep{scikit-learn} library to develop code related to classifier training.

\begin{table}[htb]
\centering
\caption{Training dataset size and class imbalance ratio for the LiTS and UMC datasets}
\resizebox{\textwidth}{!}{\begin{tabular}{ccccc}
\toprule
{} & \multicolumn{2}{c}{LiTS} & \multicolumn{2}{c}{UMC}\\
 \cmidrule(lr){2-3}
 \cmidrule(lr){4-5}
 Configuration & Training Samples & TP:FP ratio & Training Samples & TP:FP ratio\\
   \midrule
    Baseline & 1356 & 0.684 & 289 & 1.125\\
    Baseline + TTA & 1282 & 0.765 & 233 & 1.118\\
    MC-Dropout & 1267 & 0.710 & 325 & 1.321\\
    MC-Dropout + TTA & 1564 & 0.494 & 576 & 0.714\\
    Ensemble & 1332 & 0.743 & 166 & 2.388\\
    Ensemble + TTA & 1328 & 0.722 & 145 & 2.536\\
 \bottomrule
\end{tabular}}
\label{tab:samples_tab}
\end{table}

\subsubsection{Label prediction}

\begin{figure}[htb]
\centering
\begin{subfigure}[t]{0.45\linewidth}
\includegraphics[width=1.0\linewidth]{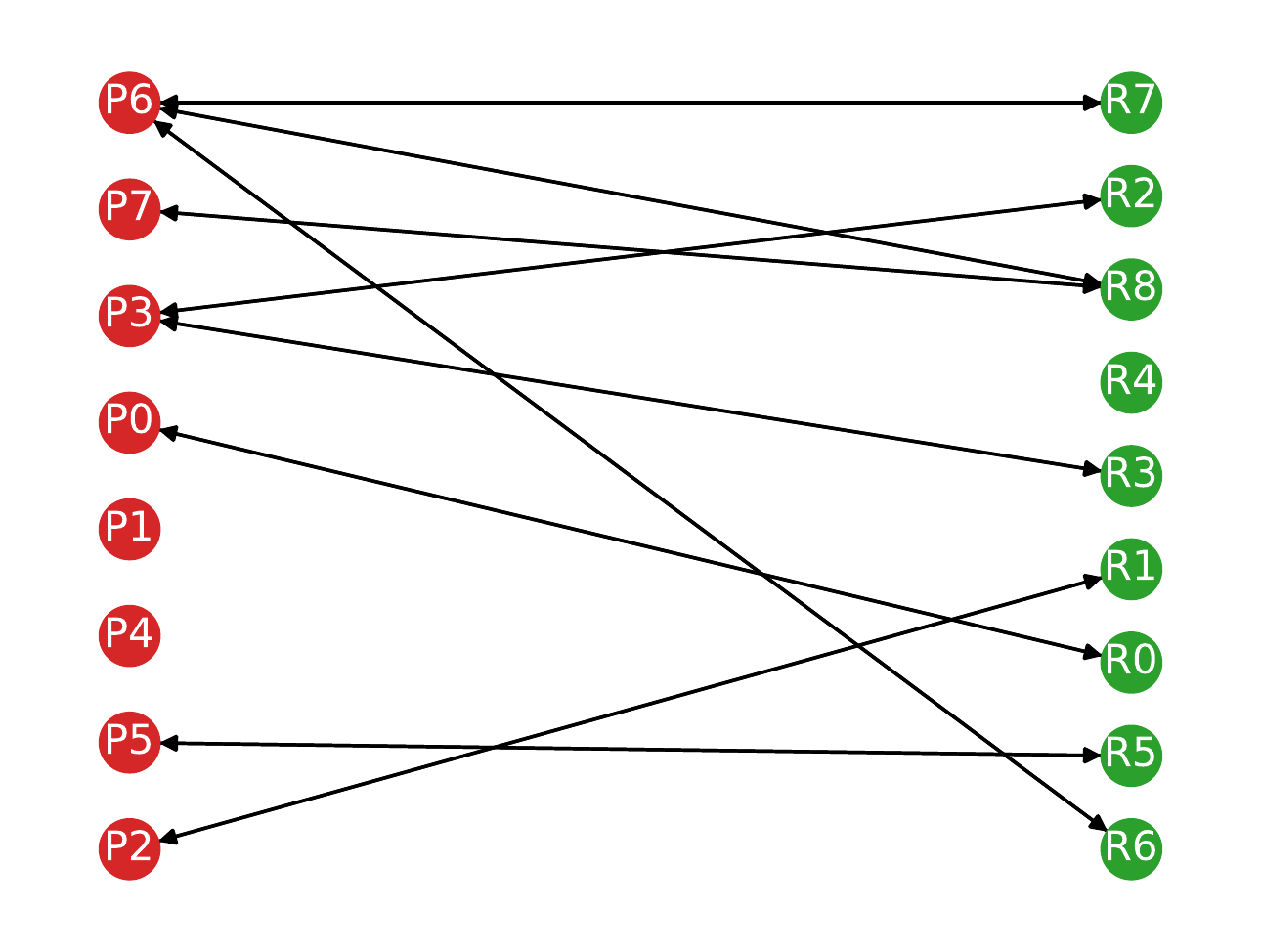}
  \caption{Lesion correspondence graph constructed from the neural network prediction}
  \label{fig:before_filtering}
\end{subfigure}
 \hspace{1em}
 \begin{subfigure}[t]{0.45\linewidth}
\includegraphics[width=1.0\linewidth]{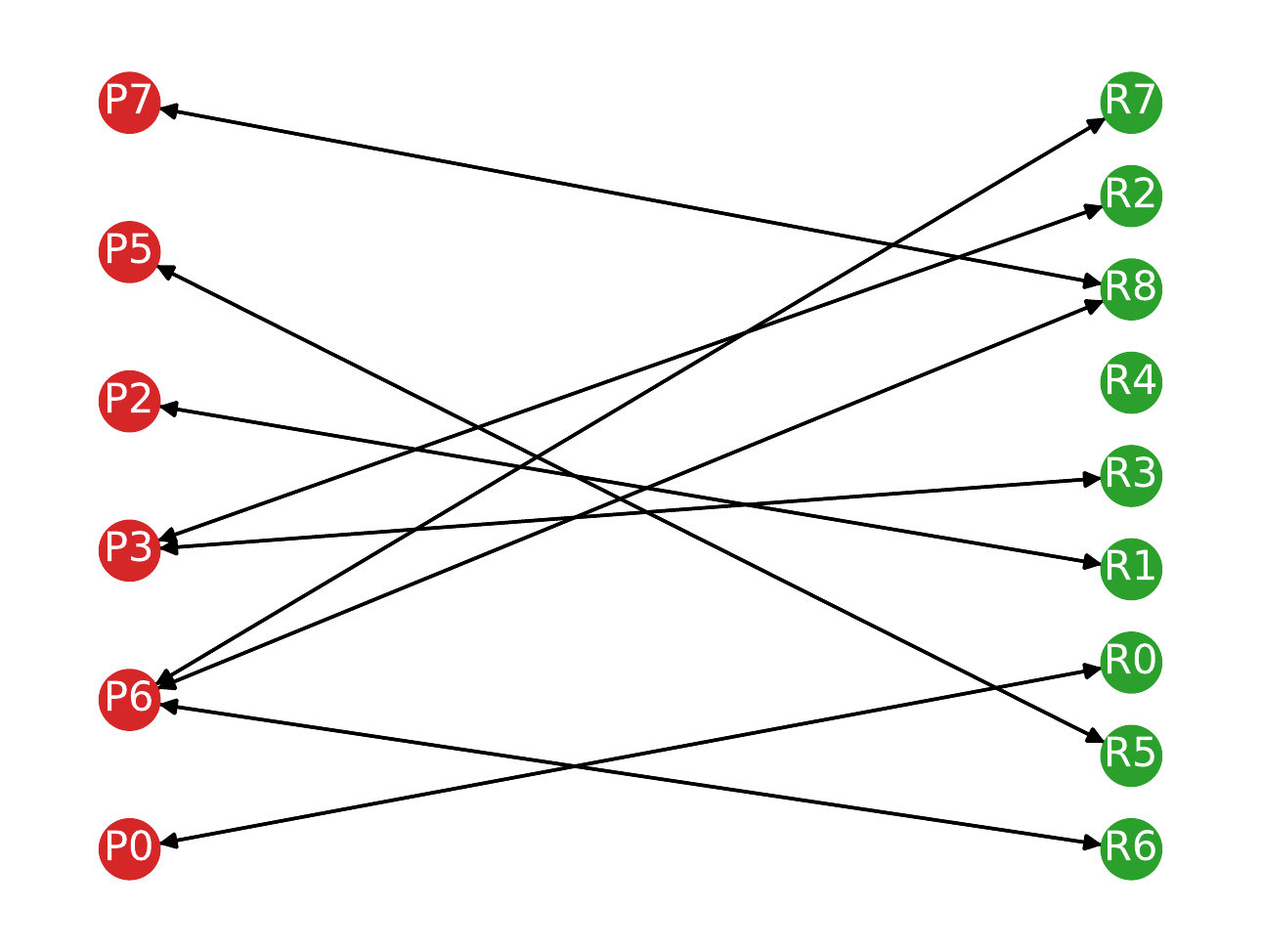}
  \caption{Lesion correspondence graph after label prediction}
  \label{fig:after_filtering}
\end{subfigure}
\caption{Modifying the lesion correspondence graph based on classifier predictions. All the false positives (detected structures P1, P4 in Figure \ref{fig:before_filtering}) are filtered out.}
\label{fig:filtering}
\end{figure}

We used the trained classifier to predict the label (true positive or false positive) of each detected lesion for each patient in the held-out test set. For each patient, the predicted labels for each lesion in the neural network prediction were then used to modify the lesion correspondence graph (Section \ref{subsec:lesion_det}). If a lesion was classified as a false positive, the corresponding node in the graph was deleted. We show this in Figure \ref{fig:filtering}.

\section{Experiments}
\label{sec:exp}
\subsection{Neural network architecture and training}
\label{sec:nn_training}
We used the 2-D U-Net~\citep{ronneberger_u-net_2015} architecture to create identical segmentation networks for both the datasets. Each convolution block consisted of a $3 \times 3$ weight kernel, instance normalization and a leaky ReLU activation function. Downsampling was performed via max-pooling and up-sampling was performed via transposed convolutions. At the output layer, a $1 \times 1$ convolution was performed, followed by the application of the softmax function to obtain per-voxel class probabilities. 

The networks for the UMC and LiTS datasets were trained for approximately $10$K (batch size = $16$) and $20$K (batch size = $8$) batch iterations based on early-stopping while monitoring the decrease in the validation loss. For both datasets, we used the ADAM~\citep{kingma_adam:_2015} optimizer with a  learning rate of $10^{-4}$ and weight decay of $5 \times 10^{-5}$. To tackle class-imbalance, we used only slices containing a lesion to train the neural network. Furthermore, we used a weighted cross-entropy as the loss function, with the per-class weights inversely proportional to the fraction of voxels belonging to that class. For both datasets, data augmentation was performed by randomly rotating the input by an angle sampled from $[-45, 45]$ degrees. 

For the MC-Dropout models, dropout layers were added at the outputs of the encoder and decoder at the lowest levels, similar to the Bayesian SegNet~\citep{kendall_bayesian_2016}. The dropout rate with the lowest validation loss was chosen. For the network trained using the UMC dataset, this was $0.3$, and for the network trained using the LiTS dataset, this was $0.5$. For test-time augmentation, we use rotations with the angles uniformly sampled from the range used during data augmentation. 

The code for the neural network training and inference was developed using the PyTorch~\citep{pytorch2019} library.

For each uncertainty estimation method, we trained the false-positive classifier and used lesions from the neural network test set to evaluate the performance. For both datasets and each uncertainty estimation method, we trained five classifiers with different random seeds to report confidence intervals on our metrics.  

\subsection{Evaluation}
\label{sec:eval}
We used the precision, recall and F1-score metrics to quantify the effect of our false-positive classification pipeline on lesion detection.

To isolate the role played by features computed from uncertainty estimates in reducing false positives, we trained the classifier using just the radiomics shape-based features. These features are computed from the binary lesion mask and, therefore, are independent of the per-voxel uncertainty estimates. Similar to \cite{chlebus_automatic_2018}, we trained the classifier using radiomics features computed from image patches corresponding to lesion segmentations.

Furthermore, to study the impact of spatial aggregation of uncertainty on false-positive classification, we compare our false positive classification pipeline to the threshold-based filtering approach proposed by \cite{nair_exploring_2020}. The classification threshold was determined using a hold-out validation set. The thresholds used were $0.9995$ and $0.9999$ for the UMC and LiTS datasets, respectively. 

Additionally, we performed \emph{cross testing} and \emph{combined testing} to investigate the generalization properties of our classifier. In cross-testing, for each uncertainty estimation method, we used the classifier  trained on one dataset to predict the type of detection in the test set of the other dataset. In combined testing, for each uncertainty estimation method, we trained a classifier by merging training sets from both datasets and used it to predict the type of detection in the test set. Both these experiments used radiomics features computed from the uncertainty maps.

We measured feature importance using the leave-one-covariate-out (LOCO)~\citep{lei_distribution-free_2018} method, where feature importance was measured directly in terms of prediction performance on unseen data.

\begin{figure*}[htb]
\centering
\begin{subfigure}[t]{1.0\linewidth}
  \centering
  \includegraphics[width=0.8\linewidth]{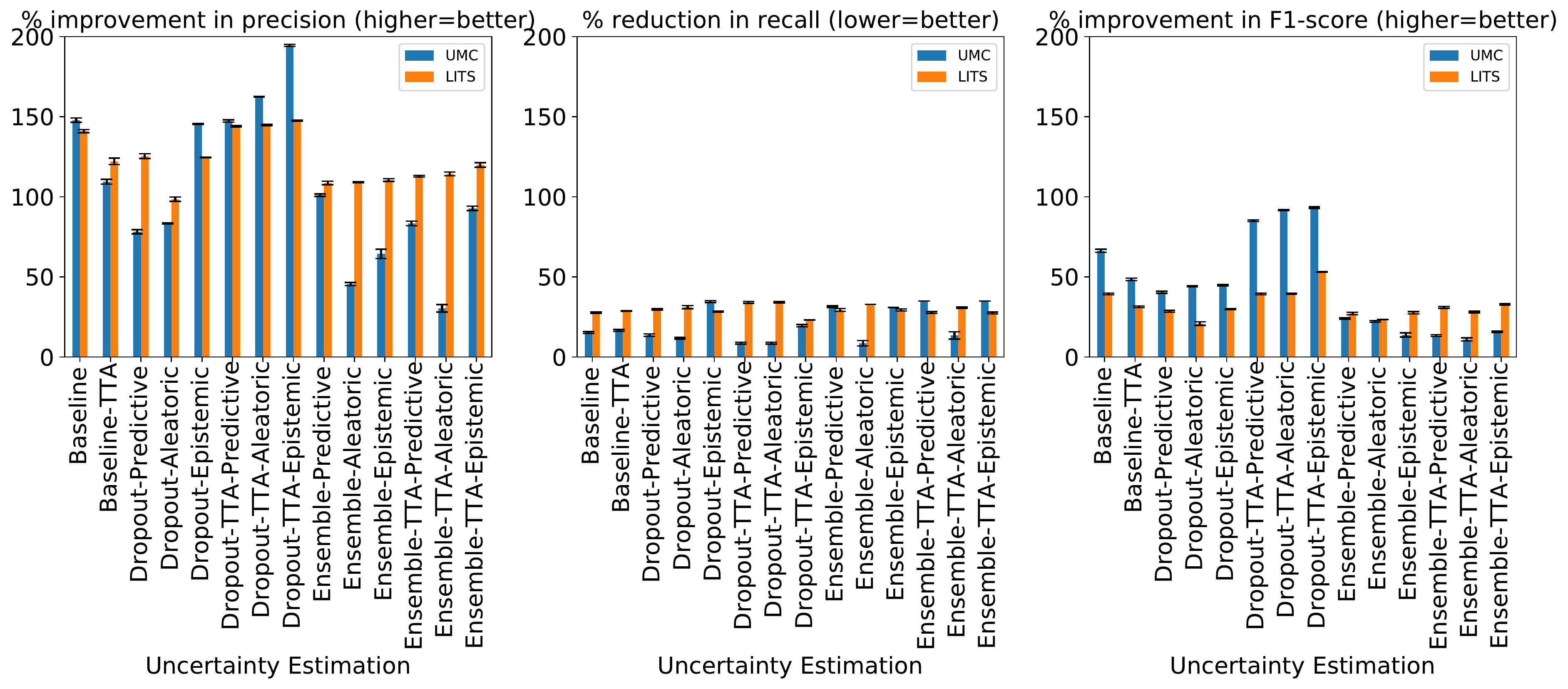}
  \caption{Radiomics features computed from uncertainty map}
  \label{fig:rel_improvement_unc}
\end{subfigure}
\begin{subfigure}[t]{1.0\linewidth}
  \centering
  \includegraphics[width=0.8\linewidth]{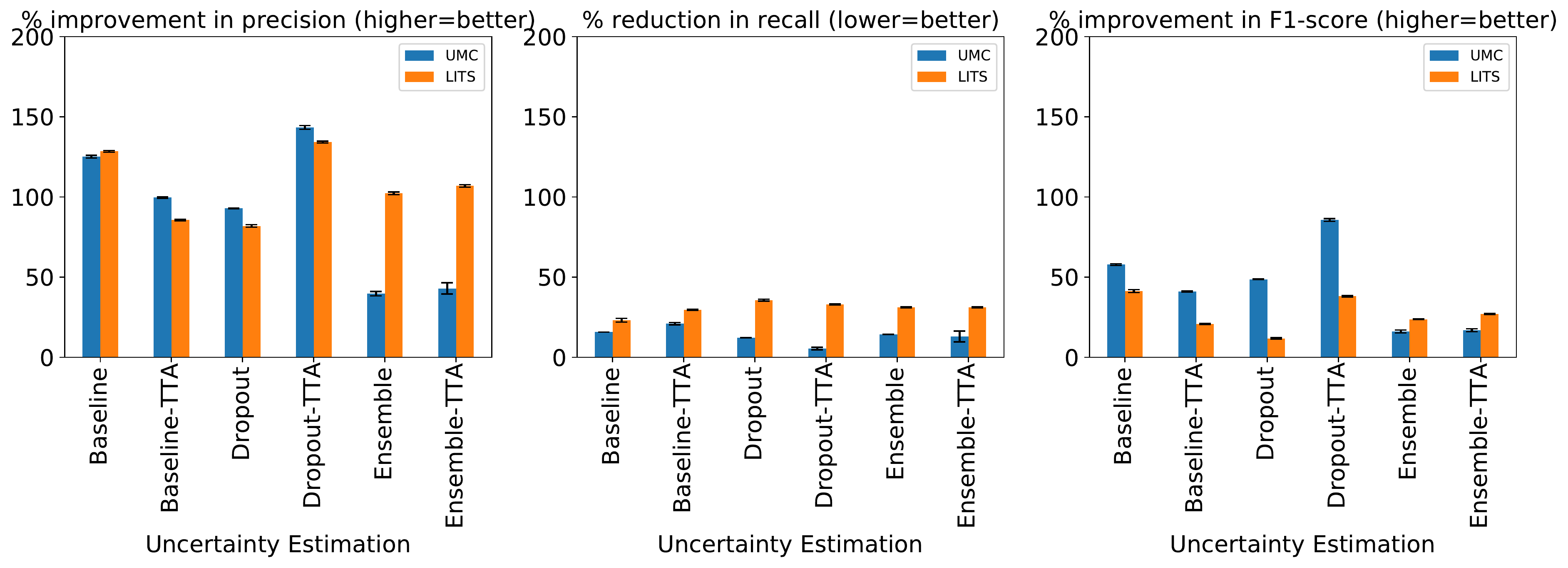}
  \caption{Radiomics features computed from image intensities}
  \label{fig:rel_improvement_intensity}
\end{subfigure}
 \begin{subfigure}[t]{1.0\linewidth}
   \centering
  \includegraphics[width=0.8\linewidth]{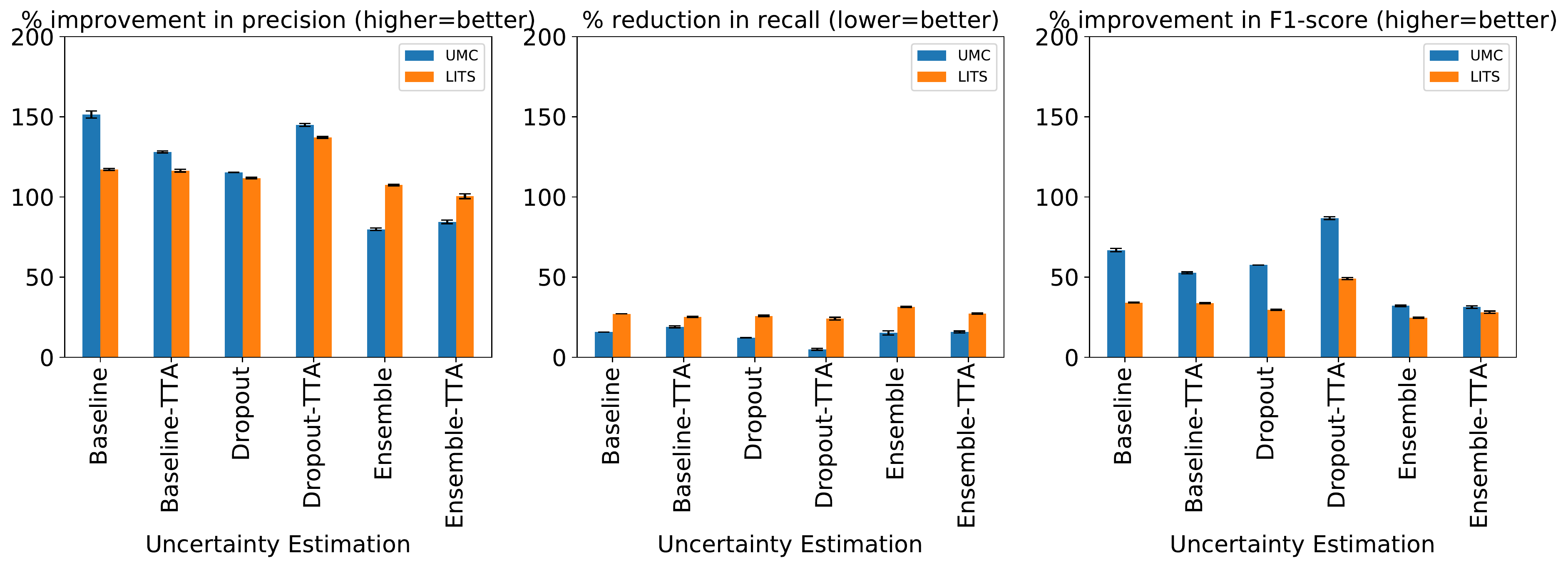}
  \caption{Radiomics features computed from binary mask}
  \label{fig:rel_improvement_shape}
\end{subfigure}
\caption{Relative change in precision, recall, and F1 metrics after false-positive classification. Figure \ref{fig:rel_improvement_unc} shows the relative change in the lesion detection metrics when radiomics features computed from uncertainty maps are used to perform false-positive classification. Figures \ref{fig:rel_improvement_intensity} and \ref{fig:rel_improvement_shape} show the relative changes in lesion detection metrics when radiomics features are computed from image patches and binary masks, respectively.}
\label{fig:rel_improvement}
\end{figure*}

\begin{figure}[t]
   \centering
  \includegraphics[width=1.0\linewidth]{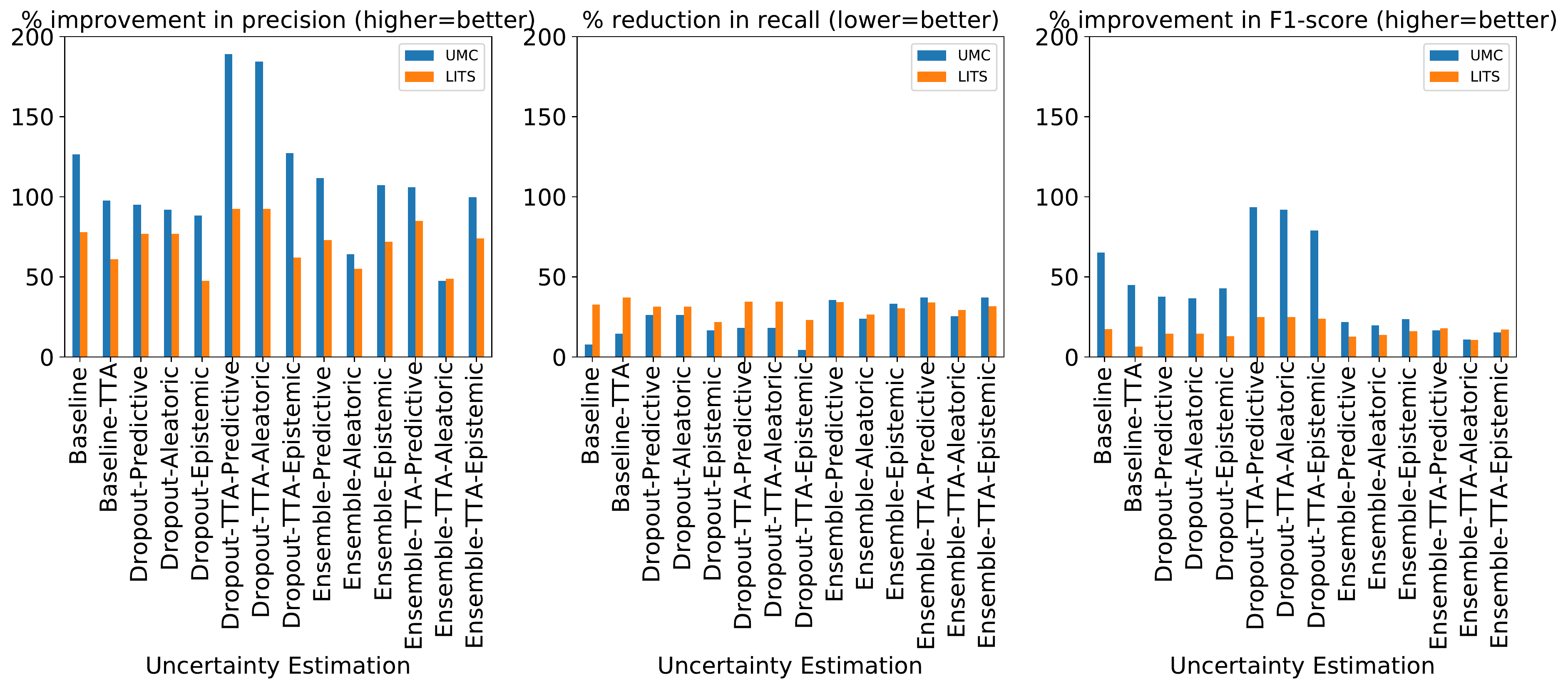}
  \caption{Relative change in the lesion detection metrics when a threshold on the log-sum of uncertainty values over the predicted lesion volume was used to filter false-positive lesions~\citep{nair_exploring_2020}}
  \label{fig:rel_improvement_logsum}
\end{figure}

\begin{table*}[htb]
\centering
\caption{Lesion detection metrics for the LiTS and UMC datasets before and after false-positive reduction}
\resizebox{\textwidth}{!}{\begin{tabular}{cccccccccccccc}
\toprule
{} & {} & \multicolumn{6}{c}{LiTS} & \multicolumn{6}{c}{UMC}\\
 \cmidrule(lr){3-8}
 \cmidrule(lr){9-14}
{} &  {} & \multicolumn{2}{c}{Precision} & \multicolumn{2}{c}{Recall} & \multicolumn{2}{c}{F1} & \multicolumn{2}{c}{Precision} & \multicolumn{2}{c}{Recall} & \multicolumn{2}{c}{F1}\\
\cmidrule(lr){3-4} \cmidrule(lr){5-6}\cmidrule(lr){7-8}\cmidrule(lr){9-10}\cmidrule(lr){11-12}\cmidrule(lr){13-14}
Configuration & Input features & Before & After & Before & After & Before & After & Before & After & Before & After & Before & After \\
\midrule
 & Predictive Uncertainty & $0.318$&$0.766 \pm 0.005$&$0.700$&$0.510 \pm 0.006$&$0.438$&$0.612 \pm 0.005$&$0.245$&$0.608 \pm 0.014$&$0.717$&$0.608 \pm 0.006$&$0.365$&$0.608 \pm 0.010$\\
 & Image & $0.318$&$0.727 \pm 0.004$&$0.700$&$0.538 \pm 0.012$&$0.438$&$0.618 \pm 0.009$&$0.245$&$0.551 \pm 0.008$&$0.717$&$0.604 \pm 0.000$&$0.365$&$0.578 \pm 0.004$\\
Baseline & Binary mask & $0.318$&$0.691 \pm 0.005$&$0.700$&$0.510 \pm 0.000$&$0.438$&$0.587 \pm 0.002$&$0.245$&$0.616 \pm 0.022$&$0.717$&$0.603 \pm 0.000$&$0.365$&$0.609 \pm 0.010$\\
\midrule
 & Predictive Uncertainty & $0.335$&$0.744 \pm 0.020$&$0.690$&$0.492 \pm 0.003$&$0.451$&$0.592 \pm 0.006$&$0.289$&$0.605 \pm 0.015$&$0.774$&$0.645 \pm 0.006$&$0.421$&$0.624 \pm 0.008$\\
 & Image & $0.335$&$0.634 \pm 0.004$&$0.690$&$0.492 \pm 0.003$&$0.451$&$0.554 \pm 0.003$&$0.289$&$0.577 \pm 0.004$&$0.774$&$0.611 \pm 0.007$&$0.421$&$0.593 \pm 0.003$\\
Baseline + TTA & Binary mask & $0.335$&$0.725 \pm 0.008$&$0.690$&$0.516 \pm 0.004$&$0.451$&$0.603 \pm 0.003$&$0.289$&$0.659 \pm 0.005$&$0.774$&$0.626 \pm 0.005$&$0.421$&$0.642 \pm 0.005$\\
\midrule
 & Predictive Uncertainty & $0.373$&$0.841 \pm 0.014$&$0.720$&$0.506 \pm 0.004$&$0.491$&$0.632 \pm 0.006$&$0.261$&$0.465 \pm 0.013$&$0.774$&$0.668 \pm 0.008$&$0.39$&$0.548 \pm 0.006$\\
 & Aleatoric Uncertainty & $0.373$&$0.740 \pm 0.014$&$0.720$&$0.496 \pm 0.01$&$0.491$&$0.594 \pm 0.011$&$0.261$&$0.479 \pm 0.004$&$0.774$&$0.683 \pm 0.006$&$0.39$&$0.563 \pm 0.004$\\
 & Epistemic Uncertainty & $0.373$&$0.838 \pm 0.001$&$0.720$&$0.516 \pm 0.004$&$0.491$&$0.639 \pm 0.003$&$0.261$&$0.641 \pm 0.003$&$0.774$&$0.506 \pm 0.006$&$0.39$&$0.565 \pm 0.005$\\
  & Image & $0.373$&$0.695 \pm 0.007$&$0.720$&$0.470 \pm 0.006$&$0.491$&$0.561 \pm 0.003$&$0.261$&$0.494 \pm 0.002$&$0.774$&$0.679 \pm 0.000$&$0.39$&$0.572 \pm 0.001$\\
MC-Dropout & Binary mask & $0.373$&$0.790 \pm 0.004$&$0.720$&$0.534 \pm 0.004$&$0.491$&$0.637 \pm 0.004$&$0.261$&$0.563 \pm 0.000$&$0.774$&$0.679 \pm 0.000$&$0.39$&$0.615 \pm 0.000$\\
\midrule
 & Predictive Uncertainty & $0.3$&$0.732 \pm 0.004$&$0.78$&$0.514 \pm 0.006$&$0.433$&$0.604 \pm 0.005$&$0.204$&$0.504 \pm 0.007$&$0.83$&$0.758 \pm 0.006$&$0.327$&$0.605 \pm 0.005$\\
 & Aleatoric Uncertainty & $0.3$&$0.734 \pm 0.005$&$0.78$&$0.514 \pm 0.004$&$0.433$&$0.605 \pm 0.004$&$0.204$&$0.535 \pm 0.002$&$0.83$&$0.758 \pm 0.006$&$0.327$&$0.627 \pm 0.003$\\
& Epistemic Uncertainty & $0.3$&$0.743 \pm 0.003$&$0.780$&$0.6 \pm 0.0$&$0.433$&$0.664 \pm 0.001$&$0.204$&$0.6 \pm 0.006$&$0.83$&$0.668 \pm 0.008$&$0.327$&$0.632 \pm 0.005$\\
& Image & $0.3$&$0.705 \pm 0.006$&$0.780$&$0.522 \pm 0.003$&$0.433$&$0.600 \pm 0.003$&$0.204$&$0.496 \pm 0.011$&$0.83$&$0.784 \pm 0.007$&$0.327$&$0.607 \pm 0.008$\\
MC-Dropout + TTA & Binary mask & $0.3$&$0.712 \pm 0.005$&$0.780$&$0.592 \pm 0.009$&$0.433$&$0.646 \pm 0.006$&$0.204$&$0.499 \pm 0.009$&$0.83$&$0.789 \pm 0.005$&$0.327$&$0.611 \pm 0.008$\\
\midrule
& Predictive Uncertainty & $0.385$&$0.804 \pm 0.011$&$0.790$&$0.558 \pm 0.009$&$0.518$&$0.659 \pm 0.008$&$0.375$&$0.754 \pm 0.008$&$0.792$&$0.543 \pm 0.006$&$0.509$&$0.632 \pm 0.005$\\
 & Aleatoric Uncertainty & $0.385$&$0.806 \pm 0.004$&$0.790$&$0.530 \pm 0.000$&$0.518$&$0.639 \pm 0.001$&$0.375$&$0.546 \pm 0.010$&$0.792$&$0.725 \pm 0.015$&$0.509$&$0.622 \pm 0.006$\\
& Epistemic Uncertainty & $0.385$&$0.811 \pm 0.008$&$0.790$&$0.558 \pm 0.006$&$0.518$&$0.661 \pm 0.007$&$0.375$&$0.616 \pm 0.029$&$0.792$&$0.547 \pm 0.000$&$0.509$&$0.579 \pm 0.013$\\
& Image & $0.385$&$0.779 \pm 0.008$&$0.790$&$0.544 \pm 0.004$&$0.518$&$0.641 \pm 0.002$&$0.375$&$0.524 \pm 0.013$&$0.792$&$0.679 \pm 0.000$&$0.509$&$0.591 \pm 0.008$\\
Ensemble & Binary mask & $0.385$&$0.799 \pm 0.004$&$0.790$&$0.542 \pm 0.003$&$0.518$&$0.646 \pm 0.004$&$0.375$&$0.674 \pm 0.007$&$0.792$&$0.672 \pm 0.013$&$0.509$&$0.673 \pm 0.004$\\
\midrule
& Predictive Uncertainty & $0.379$&$0.807 \pm 0.005$&$0.800$&$0.578 \pm 0.006$&$0.514$&$0.674 \pm 0.006$&$0.417$&$0.765 \pm 0.014$&$0.811$&$0.528 \pm 0.000$&$0.551$&$0.625 \pm 0.005$\\
 & Aleatoric Uncertainty & $0.379$&$0.813 \pm 0.012$&$0.800$&$0.554 \pm 0.004$&$0.514$&$0.659 \pm 0.006$&$0.417$&$0.544 \pm 0.022$&$0.811$&$0.702 \pm 0.023$&$0.551$&$0.612 \pm 0.010$\\
& Epistemic Uncertainty & $0.379$&$0.834 \pm 0.014$&$0.800$&$0.580 \pm 0.006$&$0.514$&$0.684 \pm 0.005$&$0.417$&$0.805 \pm 0.014$&$0.811$&$0.528 \pm 0.000$&$0.551$&$0.638 \pm 0.004$\\
& Image & $0.379$&$0.775 \pm 0.007$&$0.800$&$0.564 \pm 0.004$&$0.514$&$0.653 \pm 0.004$&$0.417$&$0.597 \pm 0.035$&$0.811$&$0.701 \pm 0.035$&$0.551$&$0.644 \pm 0.008$\\
Ensemble + TTA & Binary mask & $0.379$&$0.760 \pm 0.015$&$0.800$&$0.582 \pm 0.003$&$0.514$&$0.659 \pm 0.007$&$0.417$&$0.770 \pm 0.011$&$0.811$&$0.683 \pm 0.005$&$0.551$&$0.724 \pm 0.007$\\
\bottomrule
\end{tabular}}
\label{tab:lesion_det_met}
\end{table*}

\section{Results}
\label{sec:results}
In Figure \ref{fig:rel_improvement} we show the relative change in the lesion detection metrics of precision, recall, and F1 after false-positive classification for both datasets and all the uncertainty estimation methods using radiomics features computed from uncertainty maps (Figure \ref{fig:rel_improvement_unc}), image intensities (Figure \ref{fig:rel_improvement_intensity}), and binary masks (Figure \ref{fig:rel_improvement_shape}). We show these detection metrics (before and after classification) in Table \ref{tab:lesion_det_met}.

The relative change in lesion detection metrics for all uncertainty methods and both dataset, using the threshold-based method described in \cite{nair_exploring_2020} is shown in Figure \ref{fig:rel_improvement_logsum}.

\begin{table*}[htb]
\centering
\caption{Change in lesion detection metrics with cross-testing. The Kolmogorov-Smirnov test was used to check for statistical significance ($\alpha = 0.05$) of the change in lesion detection metrics, statistically significant changes have been marked with $*$}
\resizebox{\textwidth}{!}{\begin{tabular}{cccccccc}
\toprule
{} & {} & \multicolumn{3}{c}{LiTS} & \multicolumn{3}{c}{UMC}\\
 \cmidrule(lr){3-5}
 \cmidrule(lr){6-8}
Configuration & Uncertainty Type & Precision ($\%$ increase) & Recall ($\%$ increase) & F1 ($\%$ increase) & Precision ($\%$ increase) & Recall ($\%$ increase) & F1 ($\%$ increase) \\
\midrule
Baseline & Predictive & $-6.35^*$ & $-17.64^*$ & $-13.48^*$ & $-19.03^*$ & $11.80^*$ & $-6.09^*$\\
Baseline + TTA & Predictive &  $-15.07^*$ & $-7.31^*$ & $-10.76^*$ & $-2.72$ & $5.84^*$ & $1.24$\\
\midrule
 & Predictive &  $-30.35^*$ & $3.16^*$ & $-12.66^*$ & $27.86^*$ & $2.26^*$ & $15.65^*$\\
MC-Dropout & Aleatoric &  $-24.33^*$ & $8.87^*$ & $-7.42^*$ & $47.42^*$ & $-0.55$ & $22.97^*$\\
 & Epistemic &  $3.98$ & $-22.09^*$ & $-13.89^*$ & $-8.39^*$ & $23.13^*$ & $6.91^*$\\
\midrule
 & Predictive &  $-1.63^*$ & $-3.89^*$ & $-2.98^*$ & $-37.33^*$ & $6.97^*$ & $-24.92^*$\\
MC-Dropout + TTA & Aleatoric & $-0.92$ & $-5.44^*$ & $-3.63^*$ & $-38.77^*$ & $6.97^*$ & $-25.62^*$ \\
 & Epistemic & $1.37$ & $-25.66^*$ & $-15.68^*$ & $-52.57^*$ & $24.29^*$ & $-32.94^*$ \\
\midrule
 & Predictive & $-5.72^*$ & $2.15$ & $-1.23^*$ & $-1.44$ & $10.41^*$ & $5.12^*$ \\
Ensemble & Aleatoric & $-32.99^*$ & $29.43^*$ & $-5.57^*$ & $0.61$ & $-1.04$ & $-0.07$ \\
 & Epistemic & $-10.51^*$ & $-0.71$ & $-5.26^*$ & $13.79^*$ & $6.89^*$ & $10.11^*$ \\
\midrule
 & Predictive & $-20.45^*$ & $-10.72^*$ & $-15.07^*$ & $7.24^*$ & $7.85^*$ & $7.60^*$  \\
Ensemble + TTA & Aleatoric & $-22.14^*$ & $4.33$ & $-8.56^*$ & $4.33$ & $-3.22$ & $1.05$  \\
 & Epistemic & $-14.26^*$ & $-9.31$ & $-11.41^*$ & $0.31$ & $7.85^*$ & $4.72^*$ \\
\bottomrule
\end{tabular}}
\label{tab:lesion_det_met_cross}
\end{table*}

\begin{table*}[htb]
\centering
\caption{Change in lesion detection metrics with a classifier trained with a combined training set. The Kolmogorov-Smirnov test was used to check for statistical significance ($\alpha = 0.05$) of the change in lesion detection metrics, statistically significant changes have been marked with $*$}
\resizebox{\textwidth}{!}{\begin{tabular}{cccccccc}
\toprule
{} & {} & \multicolumn{3}{c}{LiTS} & \multicolumn{3}{c}{UMC}\\
 \cmidrule(lr){3-5}
 \cmidrule(lr){6-8}
Configuration & Uncertainty Type & Precision ($\%$ increase) & Recall ($\%$ increase) & F1 ($\%$ increase) & Precision ($\%$ increase) & Recall ($\%$ increase) & F1 ($\%$ increase) \\
\midrule
Baseline & Predictive & $3.16^*$ & $-2.76$ & $-0.49$ & $-14.26^*$ & $9.32^*$ & $-3.91^*$\\
Baseline + TTA & Predictive &  $-0.81$ & $-0.41$ & $-0.54$ & $4.53$ & $5.26^*$ & $4.88^*$\\
\midrule
 & Predictive &  $-4.77^*$ & $-4.74^*$ & $-4.74^*$ & $-5.99$ & $10.16^*$ & $0.07$\\
MC-Dropout & Aleatoric &  $8.51^*$ & $-2.82$ & $1.43$ & $-8.44^*$ & $7.73^*$ & $-2.41$\\
 & Epistemic &  $-4.25^*$ & $-4.26^*$ & $-4.26^*$ & $-30.63^*$ & $52.99^*$ & $-0.11$\\
\midrule
 & Predictive &  $1.63^*$ & $-6.61^*$ & $1.11$ & $-16.75^*$ & $7.46^*$ & $-8.52^*$\\
MC-Dropout + TTA & Aleatoric & $13.01^*$ & $-5.06^*$ & $1.63$ & $-20.47^*$ & $4.47^*$ & $-11.76^*$ \\
 & Epistemic & $-6.32^*$ & $-17.00^*$ & $12.54^*$ & $-31.84^*$ & $24.29^*$ & $-13.31^*$ \\
\midrule
 & Predictive & $2.65$ & $-1.75$ & $-0.02$ & $-6.80^*$ & $13.19^*$ & $3.84^*$ \\
Ensemble & Aleatoric & $-1.54$ & $-3.40^*$ & $-1.50^*$ & $-5.26^*$ & $2.60$ & $-2.00$ \\
 & Epistemic & $6.04^*$ & $-1.07$ & $1.69$ & $9.10^*$ & $6.20^*$ & $7.62^*$ \\
\midrule
 & Predictive & $0.40$ & $2.07$ & $1.37$ & $2.85^*$ & $5.71^*$ & $4.53^*$  \\
Ensemble + TTA & Aleatoric & $-1.41$ & $0.00$ & $-0.6$ & $-0.77$ & $-2.15$ & $-1.22$  \\
 & Epistemic & $-0.66$ & $1.37$ & $0.54$ & $-11.35^*$ & $12.14^*$ & $1.49$ \\
\bottomrule
\end{tabular}}
\label{tab:lesion_det_met_combined}
\end{table*}

In Table \ref{tab:lesion_det_met_cross} we show the changes in the lesion detection metrics (with respect to Table \ref{tab:lesion_det_met}) when MR-CT cross-testing.

In Table \ref{tab:lesion_det_met_combined} we show the changes in the lesion detection metrics (with respect to Table \ref{tab:lesion_det_met}) when combined testing was performed.

The feature importance scores for all uncertainty estimation methods have been shown in Appendix \ref{app:feat_imp_scores} (Figures \ref{fig:feat_imp_lits_all} and \ref{fig:feat_imp_umc_all}).

\section{Discussion}
\label{sec:discussion}

\begin{figure}[htb]
    \centering
    \includegraphics[width=\linewidth]{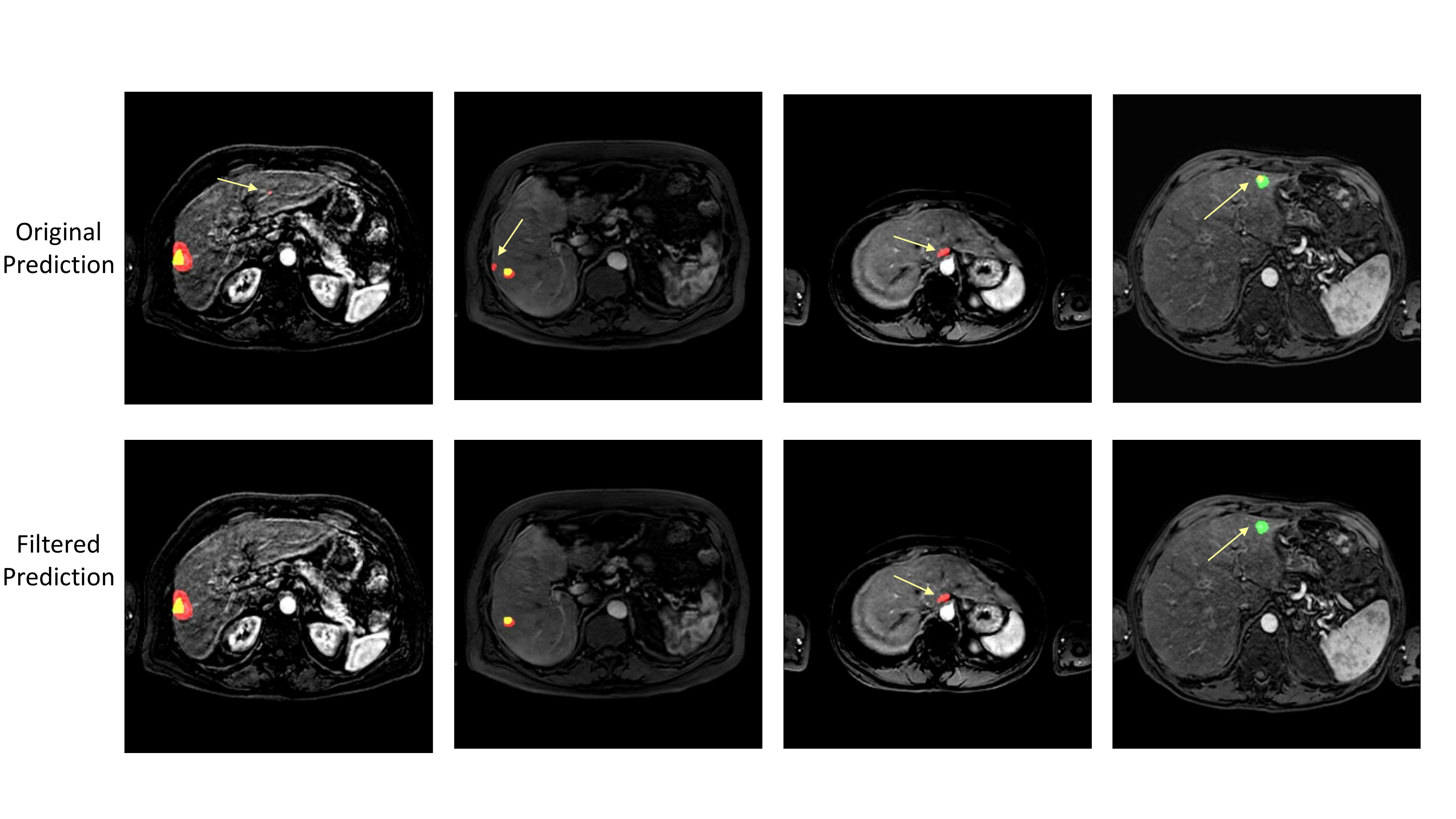}
    \caption{Influence of shape-based features on false-positive classification. The top row of the figure shows DCE-MR images from the UMC dataset overlaid with lesion segmentations computed by the segmentation network for the MC-Dropout configuration. The bottom row shows the corresponding lesion segmentation obtained after false-positive classification. The reference lesion segmentations are coloured green, the predicted lesion segmentations are red, and their overlap is yellow. The bright yellow arrow points to the predicted lesion in the original and filtered predictions. The first two columns show examples of small false positives filtered out by the classifier, while the relatively larger detections are correctly classified as true positives and retained. The third column shows an example of a slightly larger false positive misclassified as a true positive. The fourth column shows an example of a small detection overlapping with a true lesion incorrectly filtered out by the classifier.}
    \label{fig:qa}
\end{figure}

The results in Figure \ref{fig:rel_improvement} show that the chosen uncertainty estimation method, by itself, did not contribute much towards reducing false positives. However, its impact was seen on the neural network training, detection performance and feature selection for the false-positive classifier. Figures \ref{fig:rel_improvement_unc}, \ref{fig:rel_improvement_intensity}, and \ref{fig:rel_improvement_shape} show similar trends for all uncertainty estimation methods. The uncertainty estimated by any given method does not play a major role in reducing false positives, since a similar performance is observed when radiomics features from binary masks and image intensities are used to train the classifier.

Figure \ref{fig:rel_improvement} shows that the false-positive classification pipeline had a considerable impact on the performance metrics of the neural network for all uncertainty estimation methods and type of input features. While precision and F1-score improved, the extent of improvement depended on the class imbalance present in the data to train the classifier. Table \ref{tab:samples_tab} shows that the Ensemble and Ensemble+TTA configurations of the UMC dataset have a larger degree of class imbalance and a smaller training set size than the others. This was reflected in the smaller improvement in the F1-score metric for these configurations. Future work could improve upon this, by implementing strategies that make classifiers more robust  against class imbalance, like SMOTE~\citep{chawla2002} or ADASYN~\citep{he2008}. However, Figure \ref{fig:rel_improvement_shape} and Table \ref{tab:lesion_det_met}, show a better performance for the UMC Ensemble and Ensemble+TTA when radiomics features computed from binary masks were used. This shows that shape-based features are more robust to class imbalance.

As expected, there is also a slight drop in recall, due to some true positive lesions being classified as false positives and filtered out. A trade-off between precision and recall can be made, for which the optimal classifier threshold may be chosen by the end-user.

\begin{figure*}[htb]
\centering
\begin{subfigure}[t]{0.45\linewidth}
  \includegraphics[width=1.0\linewidth]{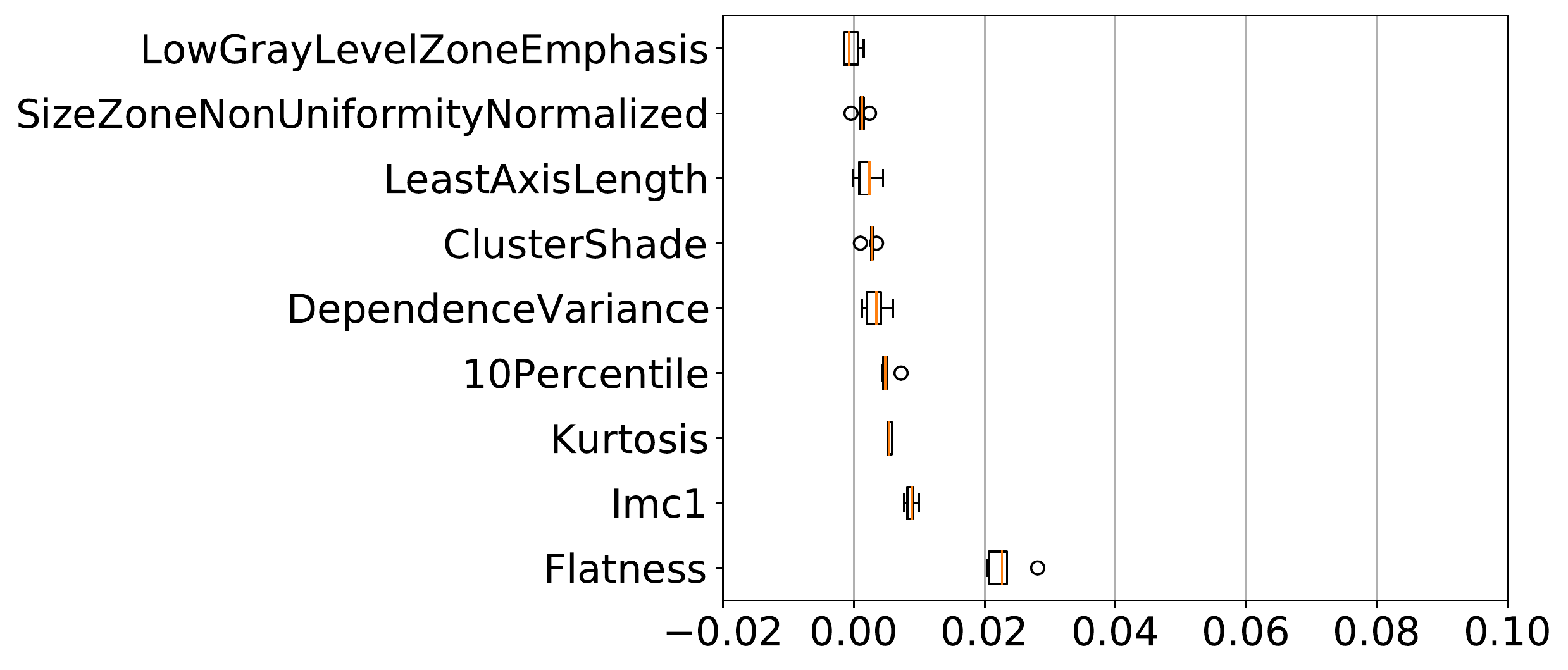}
  \caption{LOCO feature importances for LiTS}
  \label{fig:feat_imp_lits}
\end{subfigure}
\hspace{1em}
 \begin{subfigure}[t]{0.45\linewidth}
  \includegraphics[width=1.0\linewidth]{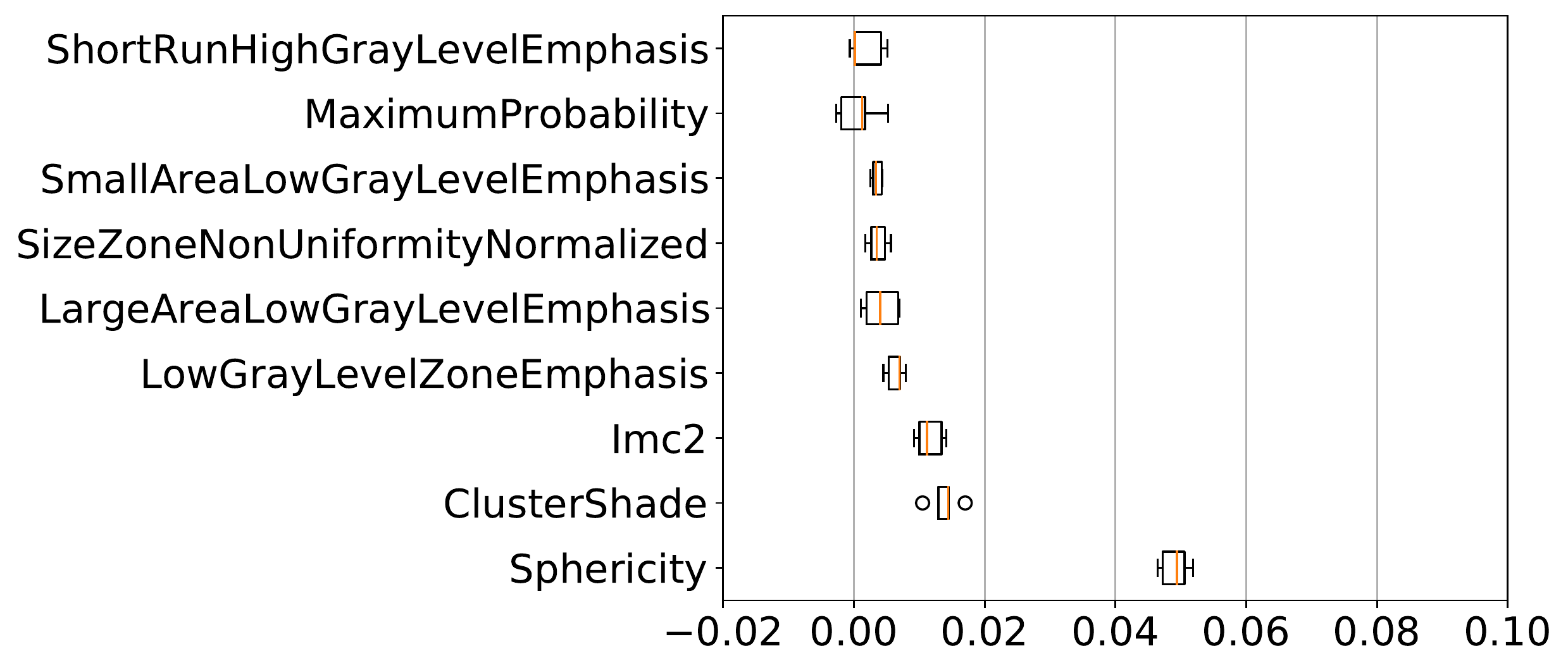}
  \caption{LOCO feature importances for UMC}
  \label{fig:feat_imp_umc}
\end{subfigure}
\hspace{1em}
\begin{subfigure}[t]{1.0\linewidth}
  \centering
  \includegraphics[width=0.6\linewidth]{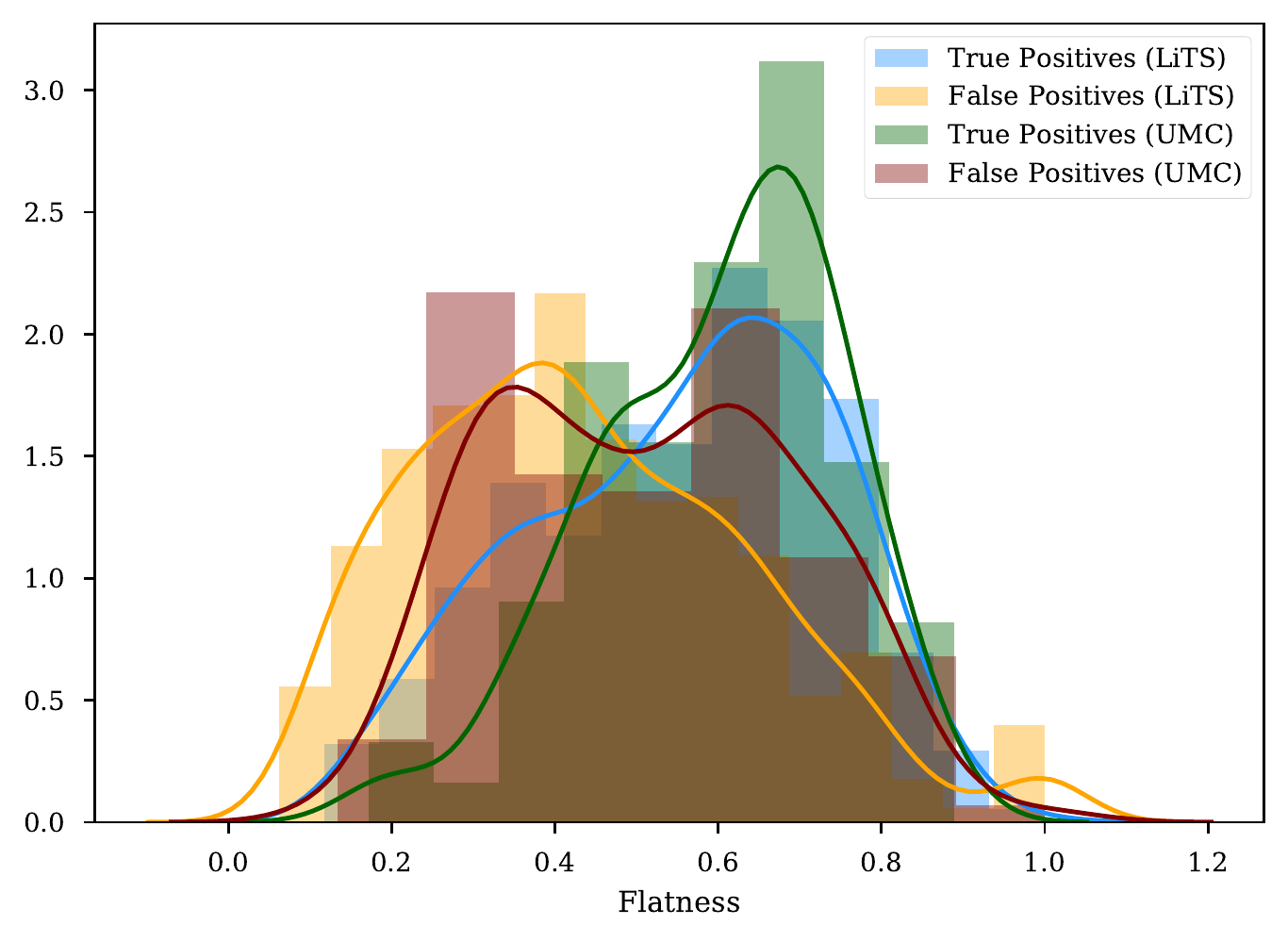}
  \caption{Distribution plot of the \emph{Flatness} feature in the training sets of LiTS and UMC data. \emph{Flatness} and \emph{Sphericity}, which are correlated features, are the most important features for the LiTS and UMC datasets respectively. We show that the different distributions of the \emph{Flatness} features in the training sets of the two datasets lead to different optimal thresholds, which cause degradation in the metrics when we perform cross-testing.}
  \label{fig:feat_dist_flatness}
\end{subfigure}
\caption{Feature importance (ascending order, top to bottom) and distributions for the \emph{Flatness} feature for the Baseline configuration}
\label{fig:feat_dist}
\end{figure*}

Across both datasets, for  almost all uncertainty estimation methods and types, the shape-based features of \emph{Sphericity} or \emph{Flatness} or \emph{Elongation} were consistently ranked highest with respect to the LOCO feature importance scores. In Figure \ref{fig:rel_improvement_shape} we see almost no difference in the relative change in the metrics when intensity and texture features are excluded from the classifier training. The trends in Figures \ref{fig:rel_improvement_unc}, \ref{fig:rel_improvement_intensity}, and \ref{fig:rel_improvement_shape} show us that features computed from the per-voxel uncertainty estimates do not play a major role in reducing false positives in this setting. In Figure \ref{fig:qa}, we show an example of how classifying a detected lesion as a false-positive depends on its size. This can be also seen in the relative improvement in the F1 score, which is almost the same for all uncertainty types for a given uncertainty estimation method.

The trends in Figure \ref{fig:rel_improvement_logsum} show that threshold-based classification using the log-sum of per-voxel uncertainty estimates is an effective method to reduce false positives. Compared to Figure \ref{fig:rel_improvement}, the trends in the relative change of precision, recall, and F1 metrics are similar for the UMC dataset, while for the LiTS dataset they are worse. The log-sum computed over a predicted lesion strongly correlates with its size. This leads to most small segmentations being classified as false positives. This trend is also reported by \cite{nair_exploring_2020}. Therefore, the log-sum aggregate is a proxy for lesion size, thereby strengthening our claim that shape-based features, and not uncertainty estimates themselves, play an important role in false-positive classification. We show evidence of this correlation in Appendix \ref{app:logsum}. The presence of smaller true positive lesions in the LiTS dataset could be a reason for the worsening of performance with respect to the false positive classification pipeline.

The results on the cross-testing (Table \ref{tab:lesion_det_met_cross}) and combined-testing (Table \ref{tab:lesion_det_met_combined}) show that, although the underlying task is the same, the trained classifier cannot be shared between datasets. This is likely because the learned features are similar, but the thresholds learned do not generalize across datasets. In Figure \ref{fig:feat_dist} we show while correlated features (\emph{Flatness} and \emph{Sphericity}) play an important role in classifying lesion predictions in each of the datasets, the classifier produces poor results when used in a \emph{cross} setting.


A limitation of our work was the inability to study the role uncertainty estimation can play in recovering false negatives. This was because the first step in the pipeline, identifying regions of interest, did not include regions for undetected lesions. Moreover, the uncertainty in the region of false negatives was low, indicating that the neural networks were incorrect with high confidence; a consequence of miscalibration. Miscalibration refers to the effect that the confidence assigned to an outcome by a classifier does not  correspond to the eventual prediction accuracy. This has been shown to occur in deep neural networks, where recent architectural advances have lead to improved classification performance but poorly modelled probability estimates~\citep{guo_calibration_2017}. In our work, this was substantiated by the fact that the false-positive classifiers did not assign a high importance to the intensity-based features computed from the uncertainty maps, but rather focussed on shape-based features computed from the binary lesion masks (Figure \ref{fig:rel_improvement_shape}, Figure \ref{fig:feat_imp_lits_all}, and Figure \ref{fig:feat_imp_umc_all}). This phenomenon was demonstrated by \cite{jungo_analyzing_2020}, who showed that voxel-wise uncertainties were insufficient for detecting segmentation failures. Improving neural network calibration would improve lesion detection metrics and might produce uncertainty estimates that are more informative for false positive reduction.

Future work could consider a number of different steps to further improve our results. Our false-positive classification pipeline\footnote{\url{https://github.com/ishaanb92/FPCPipeline}} can be used with any probabilistic classifier. Therefore, analysing the influence of uncertainty estimates computed by more recent methods~\citep{amersfoort_uncertainty_2020, liu_simple_2022} on false-positive classification is an interesting research direction. \cite{jungo_analyzing_2020} show that the calibration error of segmentation models goes down as training set size is increased. Therefore, the effect of training set size on our false-positive classification pipeline may be studied in a future work. \cite{sander_automatic_2020} train a second neural network using the uncertainty map computed by the segmentation network to correct local errors. A similar method to detect false positives is an interesting future direction because explicit feature computation is avoided. Finally, steps taken to make models robust to class-imbalance, such as selective sampling of slices or using weighted versions of loss functions during training, while improving detection metrics (especially recall), might harm calibration. Poor calibration is caused by poorly modelled probabilities, which makes it difficult to trust the uncertainty metrics computed using these probabilities. This is reflected in our results, where uncertainty estimates by themselves are not meaningful in reducing false-positive predictions. Therefore, an important direction for future research is developing segmentation models robust to class-imbalance, while still producing well-calibrated probabilities.

Our results on the LiTS and UMC datasets showed that per-voxel uncertainty estimates did not play a major role in false-positive classification. Similar to \cite{jungo_analyzing_2020}, we observed that different uncertainty estimation methods affected segmentation performance via their influence on the training dynamics and their regularization effects. Our results show that model ensembles perform the best with respect to the F1-score on both datasets.

\section{Conclusion}

We studied the efficacy of features computed from uncertainty estimates at reducing false positives  by developing a classifier-based pipeline. We found that the relative improvement in the lesion detection metrics is mainly influenced by the class imbalance in the data used to train the classifier and the distribution of various shape-based features for all the uncertainty estimation methods we studied.


\acks{This work was financially supported by the project IMPACT (Intelligence based iMprovement of Personalized treatment And Clinical workflow supporT) in the framework of the EU research programme ITEA3 (Information Technology for European Advancement). }

%
\ethics{The work follows appropriate ethical standards in conducting research and writing the manuscript, following all applicable laws and regulations regarding treatment of animals or human subjects. The UMCU Medical Ethical Committee has reviewed this study and informed consent was waived due to its retrospective nature.}

\coi{The authors declare no conflict of interest.}

\bibliography{sample}

\begin{thebibliography}{60}
\providecommand{\natexlab}[1]{#1}
\providecommand{\url}[1]{\texttt{#1}}
\expandafter\ifx\csname urlstyle\endcsname\relax
  \providecommand{\doi}[1]{doi: #1}\else
  \providecommand{\doi}{doi: \begingroup \urlstyle{rm}\Url}\fi

\bibitem[Bhat et~al.(2021)Bhat, Kuijf, Cheplygina, and Pluim]{bhat_using_2021}
Ishaan Bhat, Hugo~J. Kuijf, Veronika Cheplygina, and Josien~P.W. Pluim.
\newblock Using {Uncertainty} {Estimation} {To} {Reduce} {False} {Positives}
  {In} {Liver} {Lesion} {Detection}.
\newblock In \emph{2021 {IEEE} 18th {International} {Symposium} on {Biomedical}
  {Imaging} ({ISBI})}, pages 663--667, April 2021.
\newblock \doi{10.1109/ISBI48211.2021.9434119}.
\newblock ISSN: 1945-8452.

\bibitem[Bilic et~al.(2019)Bilic, Christ, Vorontsov, Chlebus, Chen, Dou, Fu,
  Han, Heng, Hesser, Kadoury, Konopczynski, Le, Li, Li, Lipkovà, Lowengrub,
  Meine, Moltz, Pal, Piraud, Qi, Qi, Rempfler, Roth, Schenk, Sekuboyina,
  Vorontsov, Zhou, Hülsemeyer, Beetz, Ettlinger, Gruen, Kaissis, Lohöfer,
  Braren, Holch, Hofmann, Sommer, Heinemann, Jacobs, Mamani, van Ginneken,
  Chartrand, Tang, Drozdzal, Ben-Cohen, Klang, Amitai, Konen, Greenspan,
  Moreau, Hostettler, Soler, Vivanti, Szeskin, Lev-Cohain, Sosna, Joskowicz,
  and Menze]{bilic_liver_2019}
Patrick Bilic, Patrick~Ferdinand Christ, Eugene Vorontsov, Grzegorz Chlebus,
  Hao Chen, Qi~Dou, Chi-Wing Fu, Xiao Han, Pheng-Ann Heng, Jürgen Hesser,
  Samuel Kadoury, Tomasz Konopczynski, Miao Le, Chunming Li, Xiaomeng Li, Jana
  Lipkovà, John Lowengrub, Hans Meine, Jan~Hendrik Moltz, Chris Pal, Marie
  Piraud, Xiaojuan Qi, Jin Qi, Markus Rempfler, Karsten Roth, Andrea Schenk,
  Anjany Sekuboyina, Eugene Vorontsov, Ping Zhou, Christian Hülsemeyer, Marcel
  Beetz, Florian Ettlinger, Felix Gruen, Georgios Kaissis, Fabian Lohöfer,
  Rickmer Braren, Julian Holch, Felix Hofmann, Wieland Sommer, Volker
  Heinemann, Colin Jacobs, Gabriel Efrain~Humpire Mamani, Bram van Ginneken,
  Gabriel Chartrand, An~Tang, Michal Drozdzal, Avi Ben-Cohen, Eyal Klang,
  Marianne~M. Amitai, Eli Konen, Hayit Greenspan, Johan Moreau, Alexandre
  Hostettler, Luc Soler, Refael Vivanti, Adi Szeskin, Naama Lev-Cohain, Jacob
  Sosna, Leo Joskowicz, and Bjoern~H. Menze.
\newblock The {Liver} {Tumor} {Segmentation} {Benchmark} ({LiTS}).
\newblock \emph{arXiv:1901.04056 [cs]}, January 2019.
\newblock URL \url{http://arxiv.org/abs/1901.04056}.
\newblock arXiv: 1901.04056.

\bibitem[Blundell et~al.(2015)Blundell, Cornebise, Kavukcuoglu, and
  Wierstra]{blundell_weight_2015}
Charles Blundell, Julien Cornebise, Koray Kavukcuoglu, and Daan Wierstra.
\newblock Weight uncertainty in neural networks.
\newblock In \emph{Proceedings of the 32nd International Conference on
  International Conference on Machine Learning - Volume 37}, ICML'15, page
  1613–1622. JMLR.org, 2015.

\bibitem[Breiman(1996)]{breiman_bagging_1996}
Leo Breiman.
\newblock Bagging {Predictors}.
\newblock \emph{Machine Learning}, 24\penalty0 (2):\penalty0 123--140, August
  1996.
\newblock ISSN 1573-0565.
\newblock \doi{10.1023/A:1018054314350}.
\newblock URL \url{https://doi.org/10.1023/A:1018054314350}.

\bibitem[Camarasa et~al.(2021)Camarasa, Bos, Hendrikse, Nederkoorn, Kooi,
  van~der Lugt, and de~Bruijne]{camarasa2021}
Robin Camarasa, Daniel Bos, Jeroen Hendrikse, Paul Nederkoorn, M.~Eline Kooi,
  Aad van~der Lugt, and Marleen de~Bruijne.
\newblock A quantitative comparison of epistemic uncertainty maps applied to
  multi-class segmentation.
\newblock \emph{Machine Learning for Biomedical Imaging}, 1, 2021.

\bibitem[Castro et~al.(2020)Castro, Walker, and Glocker]{castro_causality_2020}
Daniel~C. Castro, Ian Walker, and Ben Glocker.
\newblock Causality matters in medical imaging.
\newblock \emph{Nature Communications}, 11\penalty0 (1):\penalty0 3673, July
  2020.
\newblock ISSN 2041-1723.
\newblock \doi{10.1038/s41467-020-17478-w}.

\bibitem[Chawla et~al.(2002)Chawla, Bowyer, Hall, and Kegelmeyer]{chawla2002}
Nitesh~V. Chawla, Kevin~W. Bowyer, Lawrence~O. Hall, and W.~Philip Kegelmeyer.
\newblock Smote: Synthetic minority over-sampling technique.
\newblock \emph{J. Artif. Int. Res.}, 16\penalty0 (1):\penalty0 321–357, June
  2002.
\newblock ISSN 1076-9757.

\bibitem[Ching et~al.(2018)Ching, Himmelstein, Beaulieu-Jones, Kalinin, Do,
  Way, Ferrero, Agapow, Zietz, Hoffman, Xie, Rosen, Lengerich, Israeli,
  Lanchantin, Woloszynek, Carpenter, Shrikumar, Xu, Cofer, Lavender, Turaga,
  Alexandari, Lu, Harris, DeCaprio, Qi, Kundaje, Peng, Wiley, Segler, Boca,
  Swamidass, Huang, Gitter, and Greene]{ching_opportunities_2018}
Travers Ching, Daniel~S. Himmelstein, Brett~K. Beaulieu-Jones, Alexandr~A.
  Kalinin, Brian~T. Do, Gregory~P. Way, Enrico Ferrero, Paul-Michael Agapow,
  Michael Zietz, Michael~M. Hoffman, Wei Xie, Gail~L. Rosen, Benjamin~J.
  Lengerich, Johnny Israeli, Jack Lanchantin, Stephen Woloszynek, Anne~E.
  Carpenter, Avanti Shrikumar, Jinbo Xu, Evan~M. Cofer, Christopher~A.
  Lavender, Srinivas~C. Turaga, Amr~M. Alexandari, Zhiyong Lu, David~J. Harris,
  Dave DeCaprio, Yanjun Qi, Anshul Kundaje, Yifan Peng, Laura~K. Wiley, Marwin
  H.~S. Segler, Simina~M. Boca, S.~Joshua Swamidass, Austin Huang, Anthony
  Gitter, and Casey~S. Greene.
\newblock Opportunities and obstacles for deep learning in biology and
  medicine.
\newblock \emph{Journal of The Royal Society Interface}, 15\penalty0
  (141):\penalty0 20170387, April 2018.
\newblock ISSN 1742-5689, 1742-5662.
\newblock \doi{10.1098/rsif.2017.0387}.

\bibitem[Chlebus et~al.(2018)Chlebus, Schenk, Moltz, van Ginneken, Hahn, and
  Meine]{chlebus_automatic_2018}
Grzegorz Chlebus, Andrea Schenk, Jan~Hendrik Moltz, Bram van Ginneken,
  Horst~Karl Hahn, and Hans Meine.
\newblock Automatic liver tumor segmentation in {CT} with fully convolutional
  neural networks and object-based postprocessing.
\newblock \emph{Scientific Reports}, 8\penalty0 (1):\penalty0 15497, December
  2018.
\newblock ISSN 2045-2322.
\newblock \doi{10.1038/s41598-018-33860-7}.
\newblock URL \url{http://www.nature.com/articles/s41598-018-33860-7}.

\bibitem[Ciga et~al.(2021)Ciga, Xu, Nofech-Mozes, Noy, Lu, and
  Martel]{ciga_overcoming_2021}
Ozan Ciga, Tony Xu, Sharon Nofech-Mozes, Shawna Noy, Fang-I. Lu, and Anne~L.
  Martel.
\newblock Overcoming the limitations of patch-based learning to detect cancer
  in whole slide images.
\newblock \emph{Scientific Reports}, 11\penalty0 (1):\penalty0 8894, April
  2021.
\newblock ISSN 2045-2322.
\newblock \doi{10.1038/s41598-021-88494-z}.
\newblock URL \url{https://www.nature.com/articles/s41598-021-88494-z}.
\newblock Number: 1 Publisher: Nature Publishing Group.

\bibitem[Depeweg et~al.(2018)Depeweg, Hernandez-Lobato, Doshi-Velez, and
  Udluft]{depeweg_decomposition_2018}
Stefan Depeweg, Jose~Miguel Hernandez-Lobato, Finale Doshi-Velez, and Steffen
  Udluft.
\newblock Decomposition of uncertainty in bayesian deep learning for efficient
  and risk-sensitive learning.
\newblock In \emph{Proceedings of the 35th International Conference on Machine
  Learning (ICML)}, volume~80, Stockholm, Sweden, 2018.

\bibitem[DeVries and Taylor(2018)]{devries_leveraging_2018}
Terrance DeVries and Graham~W. Taylor.
\newblock Leveraging {Uncertainty} {Estimates} for {Predicting} {Segmentation}
  {Quality}.
\newblock July 2018.
\newblock URL \url{http://arxiv.org/abs/1807.00502}.
\newblock arXiv: 1807.00502.

\bibitem[Eaton-Rosen et~al.(2018)Eaton-Rosen, Bragman, Bisdas, Ourselin, and
  Cardoso]{eaton-rosen_towards_2018}
Zach Eaton-Rosen, Felix Bragman, Sotirios Bisdas, Sébastien Ourselin, and
  M.~Jorge Cardoso.
\newblock Towards {Safe} {Deep} {Learning}: {Accurately} {Quantifying}
  {Biomarker} {Uncertainty} in {Neural} {Network} {Predictions}.
\newblock In Alejandro~F. Frangi, Julia~A. Schnabel, Christos Davatzikos,
  Carlos Alberola-López, and Gabor Fichtinger, editors, \emph{Medical {Image}
  {Computing} and {Computer} {Assisted} {Intervention} – {MICCAI} 2018},
  Lecture {Notes} in {Computer} {Science}, pages 691--699, Cham, 2018. Springer
  International Publishing.
\newblock ISBN 978-3-030-00928-1.
\newblock \doi{10.1007/978-3-030-00928-1_78}.

\bibitem[Fort et~al.(2020)Fort, Hu, and Lakshminarayanan]{fort_deep_2020}
Stanislav Fort, Huiyi Hu, and Balaji Lakshminarayanan.
\newblock Deep {Ensembles}: {A} {Loss} {Landscape} {Perspective}.
\newblock \emph{arXiv:1912.02757 [cs, stat]}, June 2020.
\newblock URL \url{http://arxiv.org/abs/1912.02757}.
\newblock arXiv: 1912.02757.

\bibitem[Gal and Ghahramani(2016)]{gal_dropout_2015}
Yarin Gal and Zoubin Ghahramani.
\newblock Dropout as a bayesian approximation: Representing model uncertainty
  in deep learning.
\newblock In \emph{Proceedings of the 33nd International Conference on Machine
  Learning, {ICML} 2016}, volume~48 of \emph{{JMLR} Workshop and Conference
  Proceedings}, pages 1050--1059. JMLR.org, 2016.

\bibitem[Geurts et~al.(2006)Geurts, Ernst, and Wehenkel]{geurts2006}
Pierre Geurts, Damien Ernst, and Louis Wehenkel.
\newblock Extremely randomized trees.
\newblock \emph{Mach. Learn.}, 63\penalty0 (1):\penalty0 3–42, April 2006.
\newblock ISSN 0885-6125.
\newblock \doi{10.1007/s10994-006-6226-1}.
\newblock URL \url{https://doi.org/10.1007/s10994-006-6226-1}.

\bibitem[Graham et~al.(2019)Graham, Chen, Gamper, Dou, Heng, Snead, Tsang, and
  Rajpoot]{graham_mild-net_2019}
Simon Graham, Hao Chen, Jevgenij Gamper, Qi~Dou, Pheng-Ann Heng, David Snead,
  Yee~Wah Tsang, and Nasir Rajpoot.
\newblock {MILD}-{Net}: {Minimal} information loss dilated network for gland
  instance segmentation in colon histology images.
\newblock \emph{Medical Image Analysis}, 52:\penalty0 199--211, February 2019.
\newblock ISSN 1361-8415.
\newblock \doi{10.1016/j.media.2018.12.001}.

\bibitem[Guo et~al.(2017)Guo, Pleiss, Sun, and
  Weinberger]{guo_calibration_2017}
Chuan Guo, Geoff Pleiss, Yu~Sun, and Kilian~Q. Weinberger.
\newblock On calibration of modern neural networks.
\newblock In \emph{Proceedings of the 34th International Conference on Machine
  Learning - Volume 70}, ICML'17, page 1321–1330. JMLR.org, 2017.

\bibitem[He et~al.(2008)He, Bai, Garcia, and Li]{he2008}
Haibo He, Yang Bai, Edwardo~A. Garcia, and Shutao Li.
\newblock Adasyn: Adaptive synthetic sampling approach for imbalanced learning.
\newblock In \emph{2008 IEEE International Joint Conference on Neural Networks
  (IEEE World Congress on Computational Intelligence)}, pages 1322--1328, 2008.
\newblock \doi{10.1109/IJCNN.2008.4633969}.

\bibitem[Hoel et~al.(2020)Hoel, Tram, and Sjöberg]{hoel2020}
Carl-Johan Hoel, Tommy Tram, and Jonas Sjöberg.
\newblock Reinforcement learning with uncertainty estimation for tactical
  decision-making in intersections.
\newblock In \emph{2020 IEEE 23rd International Conference on Intelligent
  Transportation Systems (ITSC)}, pages 1--7, 2020.
\newblock \doi{10.1109/ITSC45102.2020.9294407}.

\bibitem[Jansen et~al.(2017)Jansen, Kuijf, Veldhuis, Wessels, van Leeuwen, and
  Pluim]{jansen_evaluation_2017}
M~J~A Jansen, H~J Kuijf, W~B Veldhuis, F~J Wessels, M~S van Leeuwen, and J~P~W
  Pluim.
\newblock Evaluation of motion correction for clinical dynamic contrast
  enhanced {MRI} of the liver.
\newblock \emph{Physics in Medicine \& Biology}, 62\penalty0 (19):\penalty0
  7556--7568, September 2017.
\newblock ISSN 1361-6560.
\newblock \doi{10.1088/1361-6560/aa8848}.

\bibitem[Jansen et~al.(2019)Jansen, Kuijf, Niekel, Veldhuis, Wessels,
  Viergever, and Pluim]{jansenspie}
Mariëlle J.~A. Jansen, Hugo~J. Kuijf, Maarten Niekel, Wouter~B. Veldhuis,
  Frank~J. Wessels, Max~A. Viergever, and Josien P.~W. Pluim.
\newblock {Liver segmentation and metastases detection in MR images using
  convolutional neural networks}.
\newblock \emph{Journal of Medical Imaging}, 6\penalty0 (4):\penalty0 1 -- 10,
  2019.
\newblock \doi{10.1117/1.JMI.6.4.044003}.
\newblock URL \url{https://doi.org/10.1117/1.JMI.6.4.044003}.

\bibitem[Jungo et~al.(2020)Jungo, Balsiger, and Reyes]{jungo_analyzing_2020}
Alain Jungo, Fabian Balsiger, and Mauricio Reyes.
\newblock Analyzing the {Quality} and {Challenges} of {Uncertainty}
  {Estimations} for {Brain} {Tumor} {Segmentation}.
\newblock \emph{Frontiers in Neuroscience}, 14:\penalty0 282, April 2020.
\newblock ISSN 1662-453X.
\newblock \doi{10.3389/fnins.2020.00282}.

\bibitem[Karimi et~al.(2019)Karimi, Zeng, Mathur, Avinash, Mahdavi, Spadinger,
  Abolmaesumi, and Salcudean]{karimi_accurate_2019}
Davood Karimi, Qi~Zeng, Prateek Mathur, Apeksha Avinash, Sara Mahdavi, Ingrid
  Spadinger, Purang Abolmaesumi, and Septimiu~E. Salcudean.
\newblock Accurate and robust deep learning-based segmentation of the prostate
  clinical target volume in ultrasound images.
\newblock \emph{Medical Image Analysis}, 57:\penalty0 186--196, October 2019.
\newblock ISSN 1361-8415.
\newblock \doi{10.1016/j.media.2019.07.005}.

\bibitem[Kendall et~al.(2017)Kendall, Badrinarayanan, and
  Cipolla]{kendall_bayesian_2016}
Alex Kendall, Vijay Badrinarayanan, and Roberto Cipolla.
\newblock Bayesian {SegNet}: {Model} {Uncertainty} in {Deep} {Convolutional}
  {Encoder}-{Decoder} {Architectures} for {Scene} {Understanding}.
\newblock In \emph{Procedings of the {British} {Machine} {Vision} {Conference}
  2017}, page~57, London, UK, 2017. British Machine Vision Association.
\newblock ISBN 978-1-901725-60-5.
\newblock \doi{10.5244/C.31.57}.
\newblock URL \url{http://www.bmva.org/bmvc/2017/papers/paper057/index.html}.

\bibitem[Kingma and Ba(2015)]{kingma_adam:_2015}
Diederik~P. Kingma and Jimmy Ba.
\newblock Adam: {A} {Method} for {Stochastic} {Optimization}.
\newblock In \emph{3rd International Conference on Learning Representations,
  {ICLR} 2015, San Diego, CA, USA, May 7-9, 2015, Conference Track
  Proceedings}, 2015.

\bibitem[Kingma et~al.(2015)Kingma, Salimans, and
  Welling]{kingma_variational_2015}
Diederik~P. Kingma, Tim Salimans, and Max Welling.
\newblock Variational dropout and the local reparameterization trick.
\newblock In \emph{Proceedings of the 28th International Conference on Neural
  Information Processing Systems - Volume 2}, NIPS'15, page 2575–2583,
  Cambridge, MA, USA, 2015. MIT Press.

\bibitem[Kiureghian and Ditlevsen(2009)]{kiureghian_aleatory_2009}
Armen~Der Kiureghian and Ove Ditlevsen.
\newblock Aleatory or epistemic? {Does} it matter?
\newblock \emph{Structural Safety}, 31\penalty0 (2):\penalty0 105--112, March
  2009.
\newblock ISSN 0167-4730.
\newblock \doi{10.1016/j.strusafe.2008.06.020}.

\bibitem[Klein et~al.(2010)Klein, Staring, Murphy, Viergever, and
  Pluim]{elastix2010}
S.~Klein, M.~Staring, K.~Murphy, M.A. Viergever, and J.P.W. Pluim.
\newblock Elastix : a toolbox for intensity-based medical image registration.
\newblock \emph{IEEE Transactions on Medical Imaging}, 29\penalty0
  (1):\penalty0 196--205, 2010.
\newblock ISSN 0278-0062.
\newblock \doi{10.1109/TMI.2009.2035616}.

\bibitem[Lakshminarayanan et~al.(2017)Lakshminarayanan, Pritzel, and
  Blundell]{lakshminarayanan_simple_2017}
Balaji Lakshminarayanan, Alexander Pritzel, and Charles Blundell.
\newblock Simple and scalable predictive uncertainty estimation using deep
  ensembles.
\newblock In \emph{Advances in Neural Information Processing Systems 30: Annual
  Conference on Neural Information Processing Systems 2017}, pages 6402--6413,
  2017.

\bibitem[Lei et~al.(2018)Lei, G’Sell, Rinaldo, Tibshirani, and
  Wasserman]{lei_distribution-free_2018}
Jing Lei, Max G’Sell, Alessandro Rinaldo, Ryan~J. Tibshirani, and Larry
  Wasserman.
\newblock Distribution-{Free} {Predictive} {Inference} for {Regression}.
\newblock \emph{Journal of the American Statistical Association}, 113\penalty0
  (523):\penalty0 1094--1111, July 2018.
\newblock ISSN 0162-1459, 1537-274X.
\newblock \doi{10.1080/01621459.2017.1307116}.
\newblock URL
  \url{https://www.tandfonline.com/doi/full/10.1080/01621459.2017.1307116}.

\bibitem[Leibig et~al.(2017)Leibig, Allken, Ayhan, Berens, and
  Wahl]{leibig_leveraging_2017}
Christian Leibig, Vaneeda Allken, Murat~Seçkin Ayhan, Philipp Berens, and
  Siegfried Wahl.
\newblock Leveraging uncertainty information from deep neural networks for
  disease detection.
\newblock \emph{Scientific Reports}, 7\penalty0 (1):\penalty0 17816, December
  2017.
\newblock ISSN 2045-2322.
\newblock \doi{10.1038/s41598-017-17876-z}.

\bibitem[Litjens et~al.(2017)Litjens, Kooi, Bejnordi, Setio, Ciompi,
  Ghafoorian, Laak, Ginneken, and Sánchez]{litjens_survey_2017}
Geert Litjens, Thijs Kooi, Babak~Ehteshami Bejnordi, Arnaud Arindra~Adiyoso
  Setio, Francesco Ciompi, Mohsen Ghafoorian, Jeroen A. W. M. van~der Laak,
  Bram~van Ginneken, and Clara~I. Sánchez.
\newblock A survey on deep learning in medical image analysis.
\newblock \emph{Medical Image Analysis}, 42:\penalty0 60 -- 88, 2017.
\newblock ISSN 1361-8415.
\newblock \doi{https://doi.org/10.1016/j.media.2017.07.005}.

\bibitem[Liu et~al.(2022)Liu, Padhy, Ren, Lin, Wen, Jerfel, Nado, Snoek, Tran,
  and Lakshminarayanan]{liu_simple_2022}
Jeremiah~Zhe Liu, Shreyas Padhy, Jie Ren, Zi~Lin, Yeming Wen, Ghassen Jerfel,
  Zack Nado, Jasper Snoek, Dustin Tran, and Balaji Lakshminarayanan.
\newblock A {Simple} {Approach} to {Improve} {Single}-{Model} {Deep}
  {Uncertainty} via {Distance}-{Awareness}, May 2022.
\newblock URL \url{http://arxiv.org/abs/2205.00403}.
\newblock arXiv:2205.00403 [cs, stat].

\bibitem[Loquercio et~al.(2020)Loquercio, Segu, and
  Scaramuzza]{loquercio_general_2020}
Antonio Loquercio, Mattia Segu, and Davide Scaramuzza.
\newblock A {General} {Framework} for {Uncertainty} {Estimation} in {Deep}
  {Learning}.
\newblock \emph{IEEE Robotics and Automation Letters}, 5\penalty0 (2):\penalty0
  3153--3160, April 2020.
\newblock ISSN 2377-3766, 2377-3774.
\newblock \doi{10.1109/LRA.2020.2974682}.

\bibitem[MacKay(1992)]{mackay_bayesian}
David J.~C. MacKay.
\newblock {A Practical Bayesian Framework for Backpropagation Networks}.
\newblock \emph{Neural Computation}, 4\penalty0 (3):\penalty0 448--472, 05
  1992.
\newblock ISSN 0899-7667.
\newblock \doi{10.1162/neco.1992.4.3.448}.

\bibitem[Mehrtash et~al.(2020)Mehrtash, Wells, Tempany, Abolmaesumi, and
  Kapur]{mehrtash_confidence_2019}
Alireza Mehrtash, William~M. Wells, Clare~M. Tempany, Purang Abolmaesumi, and
  Tina Kapur.
\newblock Confidence calibration and predictive uncertainty estimation for deep
  medical image segmentation.
\newblock \emph{IEEE Transactions on Medical Imaging}, page 1–1, 2020.
\newblock ISSN 1558-254X.
\newblock \doi{10.1109/tmi.2020.3006437}.
\newblock URL \url{http://dx.doi.org/10.1109/TMI.2020.3006437}.

\bibitem[Mehta et~al.(2019)Mehta, Christinck, Nair, Lemaitre, Arnold, and
  Arbel]{greenspan_propagating_2019}
Raghav Mehta, Thomas Christinck, Tanya Nair, Paul Lemaitre, Douglas Arnold, and
  Tal Arbel.
\newblock Propagating {Uncertainty} {Across} {Cascaded} {Medical} {Imaging}
  {Tasks} for {Improved} {Deep} {Learning} {Inference}.
\newblock In \emph{Uncertainty for {Safe} {Utilization} of {Machine} {Learning}
  in {Medical} {Imaging} and {Clinical} {Image}-{Based} {Procedures}}, volume
  11840, pages 23--32. Springer International Publishing, Cham, 2019.
\newblock ISBN 978-3-030-32688-3 978-3-030-32689-0.
\newblock \doi{10.1007/978-3-030-32689-0_3}.

\bibitem[Nair et~al.(2020)Nair, Precup, Arnold, and Arbel]{nair_exploring_2020}
Tanya Nair, Doina Precup, Douglas~L. Arnold, and Tal Arbel.
\newblock Exploring uncertainty measures in deep networks for {Multiple}
  sclerosis lesion detection and segmentation.
\newblock \emph{Medical Image Analysis}, 59:\penalty0 101557, 2020.
\newblock ISSN 1361-8415.
\newblock \doi{https://doi.org/10.1016/j.media.2019.101557}.
\newblock URL
  \url{https://www.sciencedirect.com/science/article/pii/S1361841519300994}.

\bibitem[Neal(1996)]{neal_bayesian}
Radford~M. Neal.
\newblock \emph{Bayesian Learning for Neural Networks}.
\newblock Springer-Verlag, Berlin, Heidelberg, 1996.
\newblock ISBN 0387947248.

\bibitem[Ng et~al.(2020)Ng, Guo, Biswas, Petersen, Piechnik, Neubauer, and
  Wright]{ng_estimating_2020}
Matthew Ng, Fumin Guo, Labonny Biswas, Steffen~E. Petersen, Stefan~K. Piechnik,
  Stefan Neubauer, and Graham Wright.
\newblock Estimating {Uncertainty} in {Neural} {Networks} for {Cardiac} {MRI}
  {Segmentation}: {A} {Benchmark} {Study}.
\newblock \emph{arXiv:2012.15772 [cs, eess]}, December 2020.
\newblock URL \url{http://arxiv.org/abs/2012.15772}.
\newblock arXiv: 2012.15772.

\bibitem[Oguz et~al.(2018)Oguz, Carass, Pham, Roy, Subbana, Calabresi,
  Yushkevich, Shinohara, and Prince]{crimi_dice_2018}
Ipek Oguz, Aaron Carass, Dzung~L. Pham, Snehashis Roy, Nagesh Subbana, Peter~A.
  Calabresi, Paul~A. Yushkevich, Russell~T. Shinohara, and Jerry~L. Prince.
\newblock Dice {Overlap} {Measures} for {Objects} of {Unknown} {Number}:
  {Application} to {Lesion} {Segmentation}.
\newblock In Alessandro Crimi, Spyridon Bakas, Hugo Kuijf, Bjoern Menze, and
  Mauricio Reyes, editors, \emph{Brainlesion: {Glioma}, {Multiple} {Sclerosis},
  {Stroke} and {Traumatic} {Brain} {Injuries}}, volume 10670, pages 3--14.
  Springer International Publishing, Cham, 2018.
\newblock ISBN 978-3-319-75237-2 978-3-319-75238-9.
\newblock \doi{10.1007/978-3-319-75238-9_1}.
\newblock Series Title: Lecture Notes in Computer Science.

\bibitem[Ovadia et~al.(2019)Ovadia, Fertig, Ren, Nado, Sculley, Nowozin,
  Dillon, Lakshminarayanan, and Snoek]{ovadia_can_2019}
Yaniv Ovadia, Emily Fertig, Jie Ren, Zachary Nado, D.~Sculley, Sebastian
  Nowozin, Joshua Dillon, Balaji Lakshminarayanan, and Jasper Snoek.
\newblock Can you trust your model's uncertainty? {Evaluating} predictive
  uncertainty under dataset shift.
\newblock In \emph{Advances in {Neural} {Information} {Processing} {Systems}
  32}, pages 13991--14002. Curran Associates, Inc., 2019.

\bibitem[Paszke et~al.(2019)Paszke, Gross, Massa, Lerer, Bradbury, Chanan,
  Killeen, Lin, Gimelshein, Antiga, Desmaison, Kopf, Yang, DeVito, Raison,
  Tejani, Chilamkurthy, Steiner, Fang, Bai, and Chintala]{pytorch2019}
Adam Paszke, Sam Gross, Francisco Massa, Adam Lerer, James Bradbury, Gregory
  Chanan, Trevor Killeen, Zeming Lin, Natalia Gimelshein, Luca Antiga, Alban
  Desmaison, Andreas Kopf, Edward Yang, Zachary DeVito, Martin Raison, Alykhan
  Tejani, Sasank Chilamkurthy, Benoit Steiner, Lu~Fang, Junjie Bai, and Soumith
  Chintala.
\newblock Pytorch: An imperative style, high-performance deep learning library.
\newblock In H.~Wallach, H.~Larochelle, A.~Beygelzimer, F.~d\textquotesingle
  Alch\'{e}-Buc, E.~Fox, and R.~Garnett, editors, \emph{Advances in Neural
  Information Processing Systems 32}, pages 8024--8035. Curran Associates,
  Inc., 2019.

\bibitem[Pedregosa et~al.(2011)Pedregosa, Varoquaux, Gramfort, Michel, Thirion,
  Grisel, Blondel, Prettenhofer, Weiss, Dubourg, Vanderplas, Passos,
  Cournapeau, Brucher, Perrot, and Duchesnay]{scikit-learn}
F.~Pedregosa, G.~Varoquaux, A.~Gramfort, V.~Michel, B.~Thirion, O.~Grisel,
  M.~Blondel, P.~Prettenhofer, R.~Weiss, V.~Dubourg, J.~Vanderplas, A.~Passos,
  D.~Cournapeau, M.~Brucher, M.~Perrot, and E.~Duchesnay.
\newblock Scikit-learn: Machine learning in {P}ython.
\newblock \emph{Journal of Machine Learning Research}, 12:\penalty0 2825--2830,
  2011.

\bibitem[Ronneberger et~al.(2015)Ronneberger, Fischer, and
  Brox]{ronneberger_u-net_2015}
Olaf Ronneberger, Philipp Fischer, and Thomas Brox.
\newblock U-{Net}: {Convolutional} {Networks} for {Biomedical} {Image}
  {Segmentation}.
\newblock In \emph{Medical {Image} {Computing} and {Computer}-{Assisted}
  {Intervention} – {MICCAI} 2015}, pages 234--241, Cham, 2015. Springer
  International Publishing.
\newblock ISBN 978-3-319-24574-4.

\bibitem[Roy et~al.(2018)Roy, Conjeti, Navab, and Wachinger]{roy_inherent_2018}
Abhijit~Guha Roy, Sailesh Conjeti, Nassir Navab, and Christian Wachinger.
\newblock Inherent {Brain} {Segmentation} {Quality} {Control} from {Fully}
  {ConvNet} {Monte} {Carlo} {Sampling}.
\newblock In \emph{Medical {Image} {Computing} and {Computer} {Assisted}
  {Intervention} – {MICCAI} 2018}, pages 664--672, Cham, 2018. Springer
  International Publishing.
\newblock ISBN 978-3-030-00928-1.

\bibitem[Sander et~al.(2019)Sander, de~Vos, Wolterink, and
  Išgum]{sander_towards_2019}
Jörg Sander, Bob~D. de~Vos, Jelmer~M. Wolterink, and Ivana Išgum.
\newblock {Towards increased trustworthiness of deep learning segmentation
  methods on cardiac MRI}.
\newblock In \emph{Medical Imaging 2019: Image Processing}, volume 10949, pages
  324 -- 330. International Society for Optics and Photonics, SPIE, 2019.
\newblock \doi{10.1117/12.2511699}.
\newblock URL \url{https://doi.org/10.1117/12.2511699}.

\bibitem[Sander et~al.(2020)Sander, de~Vos, and Išgum]{sander_automatic_2020}
Jörg Sander, Bob~D. de~Vos, and Ivana Išgum.
\newblock Automatic segmentation with detection of local segmentation failures
  in cardiac {MRI}.
\newblock \emph{Scientific Reports}, 10\penalty0 (1):\penalty0 21769, December
  2020.
\newblock ISSN 2045-2322.
\newblock \doi{10.1038/s41598-020-77733-4}.
\newblock Number: 1 Publisher: Nature Publishing Group.

\bibitem[Sedai et~al.(2018)Sedai, Antony, Mahapatra, and
  Garnavi]{stoyanov_joint_2018}
Suman Sedai, Bhavna Antony, Dwarikanath Mahapatra, and Rahil Garnavi.
\newblock Joint {Segmentation} and {Uncertainty} {Visualization} of {Retinal}
  {Layers} in {Optical} {Coherence} {Tomography} {Images} {Using} {Bayesian}
  {Deep} {Learning}.
\newblock In Danail Stoyanov, Zeike Taylor, Francesco Ciompi, Yanwu Xu, Anne
  Martel, Lena Maier-Hein, Nasir Rajpoot, Jeroen van~der Laak, Mitko Veta,
  Stephen McKenna, David Snead, Emanuele Trucco, Mona~K. Garvin, Xin~Jan Chen,
  and Hrvoje Bogunovic, editors, \emph{Computational {Pathology} and
  {Ophthalmic} {Medical} {Image} {Analysis}}, volume 11039, pages 219--227.
  Springer International Publishing, Cham, 2018.
\newblock ISBN 978-3-030-00948-9 978-3-030-00949-6.
\newblock \doi{10.1007/978-3-030-00949-6_26}.
\newblock Series Title: Lecture Notes in Computer Science.

\bibitem[Seeböck et~al.(2020)Seeböck, Orlando, Schlegl, Waldstein,
  Bogunović, Klimscha, Langs, and Schmidt-Erfurth]{seebock_exploiting_2020}
Philipp Seeböck, José~Ignacio Orlando, Thomas Schlegl, Sebastian~M.
  Waldstein, Hrvoje Bogunović, Sophie Klimscha, Georg Langs, and Ursula
  Schmidt-Erfurth.
\newblock Exploiting {Epistemic} {Uncertainty} of {Anatomy} {Segmentation} for
  {Anomaly} {Detection} in {Retinal} {OCT}.
\newblock \emph{IEEE Transactions on Medical Imaging}, 39\penalty0
  (1):\penalty0 87--98, January 2020.
\newblock ISSN 0278-0062, 1558-254X.
\newblock \doi{10.1109/TMI.2019.2919951}.
\newblock URL \url{http://arxiv.org/abs/1905.12806}.
\newblock arXiv: 1905.12806.

\bibitem[{Smith} and {Gal}(2018)]{smith_understanding_2018}
L.~{Smith} and Y.~{Gal}.
\newblock {Understanding Measures of Uncertainty for Adversarial Example
  Detection}.
\newblock In \emph{UAI}, 2018.

\bibitem[Srivastava et~al.(2014)Srivastava, Hinton, Krizhevsky, Sutskever, and
  Salakhutdinov]{srivastava_dropout_2014}
Nitish Srivastava, Geoffrey Hinton, Alex Krizhevsky, Ilya Sutskever, and Ruslan
  Salakhutdinov.
\newblock Dropout: {A} {Simple} {Way} to {Prevent} {Neural} {Networks} from
  {Overfitting}.
\newblock \emph{Journal of Machine Learning Research}, 15\penalty0
  (56):\penalty0 1929--1958, 2014.

\bibitem[Tao et~al.(2019)Tao, Chen, Han, Peng, Li, Hua, and Lin]{tao_new_2019}
Chao Tao, Ke~Chen, Lin Han, Yulan Peng, Cheng Li, Zhan Hua, and Jiangli Lin.
\newblock New one-step model of breast tumor locating based on deep learning.
\newblock \emph{Journal of X-Ray Science and Technology}, 27\penalty0
  (5):\penalty0 839--856, 2019.
\newblock ISSN 1095-9114.
\newblock \doi{10.3233/XST-190548}.

\bibitem[Van~Amersfoort et~al.(2020)Van~Amersfoort, Smith, Teh, and
  Gal]{amersfoort_uncertainty_2020}
Joost Van~Amersfoort, Lewis Smith, Yee~Whye Teh, and Yarin Gal.
\newblock Uncertainty {Estimation} {Using} a {Single} {Deep} {Deterministic}
  {Neural} {Network}.
\newblock In \emph{Proceedings of the 37th {International} {Conference} on
  {Machine} {Learning}}, pages 9690--9700. PMLR, November 2020.
\newblock URL \url{https://proceedings.mlr.press/v119/van-amersfoort20a.html}.
\newblock ISSN: 2640-3498.

\bibitem[Van~Griethuysen et~al.(2017)Van~Griethuysen, Fedorov, Parmar, Hosny,
  Aucoin, Narayan, Beets-Tan, Fillion-Robin, Pieper, and
  Aerts]{van_griethuysen_computational_2017}
Joost~J.M. Van~Griethuysen, Andriy Fedorov, Chintan Parmar, Ahmed Hosny, Nicole
  Aucoin, Vivek Narayan, Regina~G.H. Beets-Tan, Jean-Christophe Fillion-Robin,
  Steve Pieper, and Hugo~J.W.L. Aerts.
\newblock Computational {Radiomics} {System} to {Decode} the {Radiographic}
  {Phenotype}.
\newblock \emph{Cancer Research}, 77\penalty0 (21):\penalty0 e104--e107, 2017.
\newblock ISSN 0008-5472.
\newblock \doi{10.1158/0008-5472.CAN-17-0339}.
\newblock Publisher: American Association for Cancer Research \_eprint:
  https://cancerres.aacrjournals.org/content/77/21/e104.full.pdf.

\bibitem[Van~Timmeren et~al.(2020)Van~Timmeren, Cester, Tanadini-Lang, Alkadhi,
  and Baessler]{van_timmeren_radiomics_2020}
Janita~E. Van~Timmeren, Davide Cester, Stephanie Tanadini-Lang, Hatem Alkadhi,
  and Bettina Baessler.
\newblock Radiomics in medical imaging—“how-to” guide and critical
  reflection.
\newblock \emph{Insights into Imaging}, 11\penalty0 (1):\penalty0 91, December
  2020.
\newblock ISSN 1869-4101.
\newblock \doi{10.1186/s13244-020-00887-2}.
\newblock URL
  \url{https://insightsimaging.springeropen.com/articles/10.1186/s13244-020-00887-2}.

\bibitem[Wang et~al.(2019)Wang, Li, Aertsen, Deprest, Ourselin, and
  Vercauteren]{wang_aleatoric_2019}
Guotai Wang, Wenqi Li, Michael Aertsen, Jan Deprest, Sébastien Ourselin, and
  Tom Vercauteren.
\newblock Aleatoric uncertainty estimation with test-time augmentation for
  medical image segmentation with convolutional neural networks.
\newblock \emph{Neurocomputing}, 338:\penalty0 34 -- 45, 2019.
\newblock ISSN 0925-2312.
\newblock \doi{https://doi.org/10.1016/j.neucom.2019.01.103}.

\bibitem[Zhou et~al.(2021)Zhou, Greenspan, Davatzikos, Duncan, Van~Ginneken,
  Madabhushi, Prince, Rueckert, and Summers]{zhou2021_review}
S.~Kevin Zhou, Hayit Greenspan, Christos Davatzikos, James~S. Duncan, Bram
  Van~Ginneken, Anant Madabhushi, Jerry~L. Prince, Daniel Rueckert, and
  Ronald~M. Summers.
\newblock A review of deep learning in medical imaging: Imaging traits,
  technology trends, case studies with progress highlights, and future
  promises.
\newblock \emph{Proceedings of the IEEE}, 109\penalty0 (5):\penalty0 820--838,
  2021.
\newblock \doi{10.1109/JPROC.2021.3054390}.

\bibitem[Zwanenburg et~al.(2020)Zwanenburg, Leger, Vallières, and
  Löck]{zwanenburg_image_2020}
Alex Zwanenburg, Stefan Leger, Martin Vallières, and Steffen Löck.
\newblock Image biomarker standardisation initiative.
\newblock \emph{Radiology}, 295\penalty0 (2):\penalty0 328--338, May 2020.
\newblock ISSN 0033-8419, 1527-1315.
\newblock \doi{10.1148/radiol.2020191145}.
\newblock URL \url{http://arxiv.org/abs/1612.07003}.
\newblock arXiv: 1612.07003.

\end{thebibliography}

\newpage
\appendix 
\section{Classifier Hyperparameters}
\label{app:class_hyperparams}
\begin{table}[htb]
\centering
\resizebox{\textwidth}{!}{\begin{tabular}{cccccccccc}
\toprule
{} & {} & \multicolumn{4}{c}{LiTS} & \multicolumn{4}{c}{UMC}\\
 \cmidrule(lr){3-6}
 \cmidrule(lr){7-10}
 Configuration & Input features & Number of trees & Minimum samples for splitting & Minimum samples per leaf & Splitting criterion & Number of trees & Minimum samples for splitting & Minimum samples per leaf & Splitting criterion\\
   \midrule
    & Predictive Uncertainty & 250 & 10 & 1 & Entropy & 250 & 4 & 1 & Entropy\\
    & Image & 1000 & 2 & 4 & Gini & 750 & 12 & 1 & Gini \\
Baseline    & Binary mask & 750 & 10 & 4 & Gini & 250 & 2 & 8 & Gini \\
\midrule
    & Predictive Uncertainty & 250 & 10 & 4 & Gini & 750 & 12 & 1 & Entropy\\
    & Image & 500 & 12 & 1 & Entropy & 250 & 4 & 1 & Gini \\
Baseline + TTA    & Binary mask & 1000 & 2 & 8 & Entropy & 250 & 12 & 1 & Entropy \\
\midrule
 & Predictive Uncertainty & 500 & 10 & 2 & Entropy & 250 & 10 & 1 & Entropy\\
 & Aleatoric Uncertainty & 250 & 2 & 4 & Entropy & 250 & 2 & 4 & Gini \\
 & Epistemic Uncertainty & 250 & 8 & 2 & Gini & 750 & 2 & 1 & Entropy \\
 & Image & 250 & 12 & 2 & Entropy & 1000 & 10 & 1 & Gini\\
Dropout & Binary mask & 750 & 10 & 4 & Entropy & 250 & 2 & 1 & Entropy \\
\midrule
 & Predictive Uncertainty & 750 & 12 & 4 & Entropy & 750 & 8 & 1 & Entropy\\
 & Aleatoric Uncertainty & 1000 & 2 &  4 & Entropy & 500 & 8 & 2 & Entropy \\
 & Epistemic Uncertainty & 250 & 8 & 2 & Entropy & 250 & 12 & 2 & Entropy \\
 & Image & 250 & 12 & 1 & Entropy & 250 & 2 & 2 & Entropy \\
Dropout + TTA & Binary mask & 250 & 12 & 4 & Entropy & 250 & 12 & 2 & Entropy \\
\midrule
 & Predictive Uncertainty & 250 & 12 & 1 & Entropy & 500 & 4 & 1 & Gini \\
 & Aleatoric Uncertainty & 250 & 8 & 2 & Entropy & 750 & 2 & 1 & Entropy \\
 & Epistemic Uncertainty & 500 & 2 & 4 & Gini & 250 & 2 & 1 & Gini \\
 & Image & 250 & 8 & 2 & Gini & 1000 & 4 & 1 & Entropy \\
Ensemble & Binary mask & 750 & 10 & 4 & Entropy & 250 & 12 & 4 & Entropy \\
\midrule
 & Predictive Uncertainty & 250 & 12 & 1 & Gini & 250 & 10 & 4 & Gini \\
 & Aleatoric Uncertainty & 500 & 12 & 2 & Entropy & 500 & 4 & 1 & Entropy \\
 & Epistemic Uncertainty & 500 & 10 & 2 & Entropy & 750 & 2 & 10 & Entropy \\
 & Image & 250 & 2 & 4 & Entropy & 750 & 2 & 1 & Entropy \\
Ensemble + TTA & Binary mask & 750 & 10 & 2 & Gini & 1000 & 2 & 4 & Gini \\
 \bottomrule
\end{tabular}}
\caption{Classifier hyperparameters for the LiTS and UMC datasets}
\label{tab:class_hyperparams}
\end{table}

\section{Feature Importance scores}
\label{app:feat_imp_scores}
\begin{figure}
    \centering
    \includegraphics[width=0.55\textwidth]{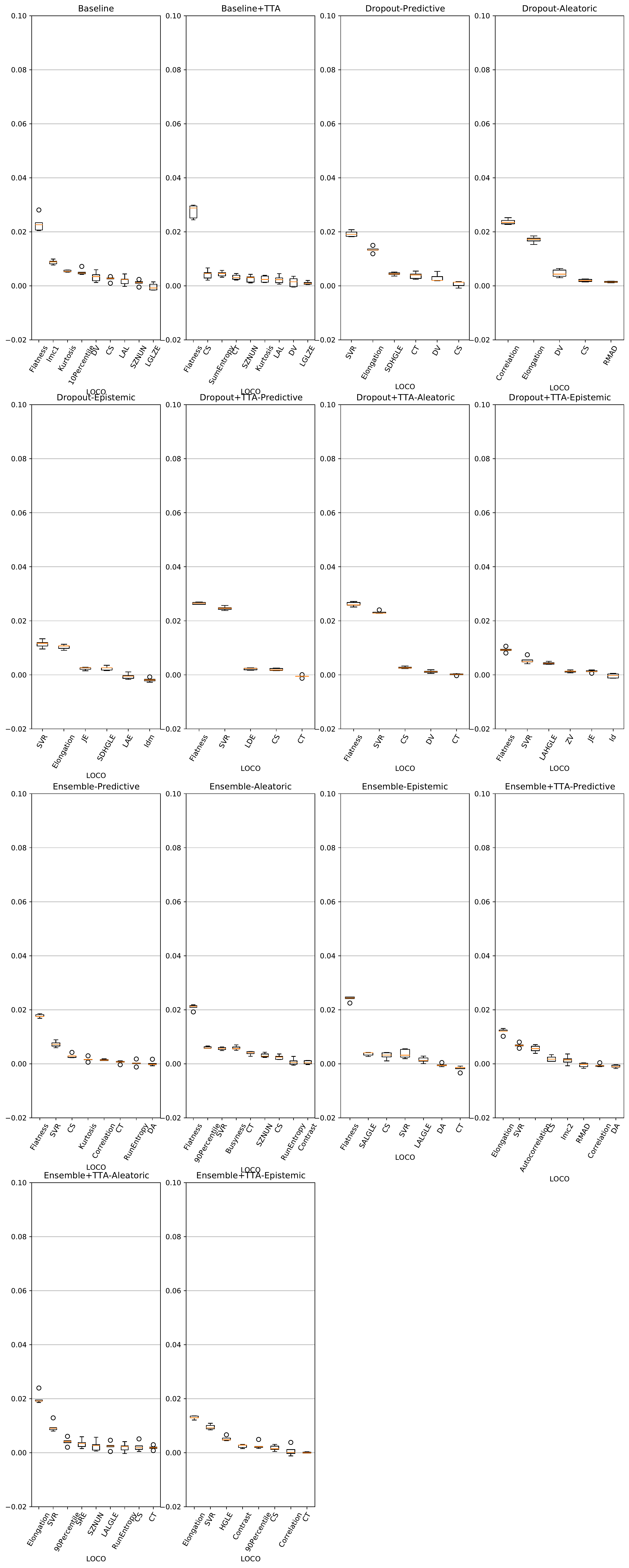}
    \caption{Feature importance scores (LOCO) for LiTS classifiers}
    \label{fig:feat_imp_lits_all}
\end{figure}

\begin{figure}
    \centering
    \includegraphics[width=0.55\textwidth]{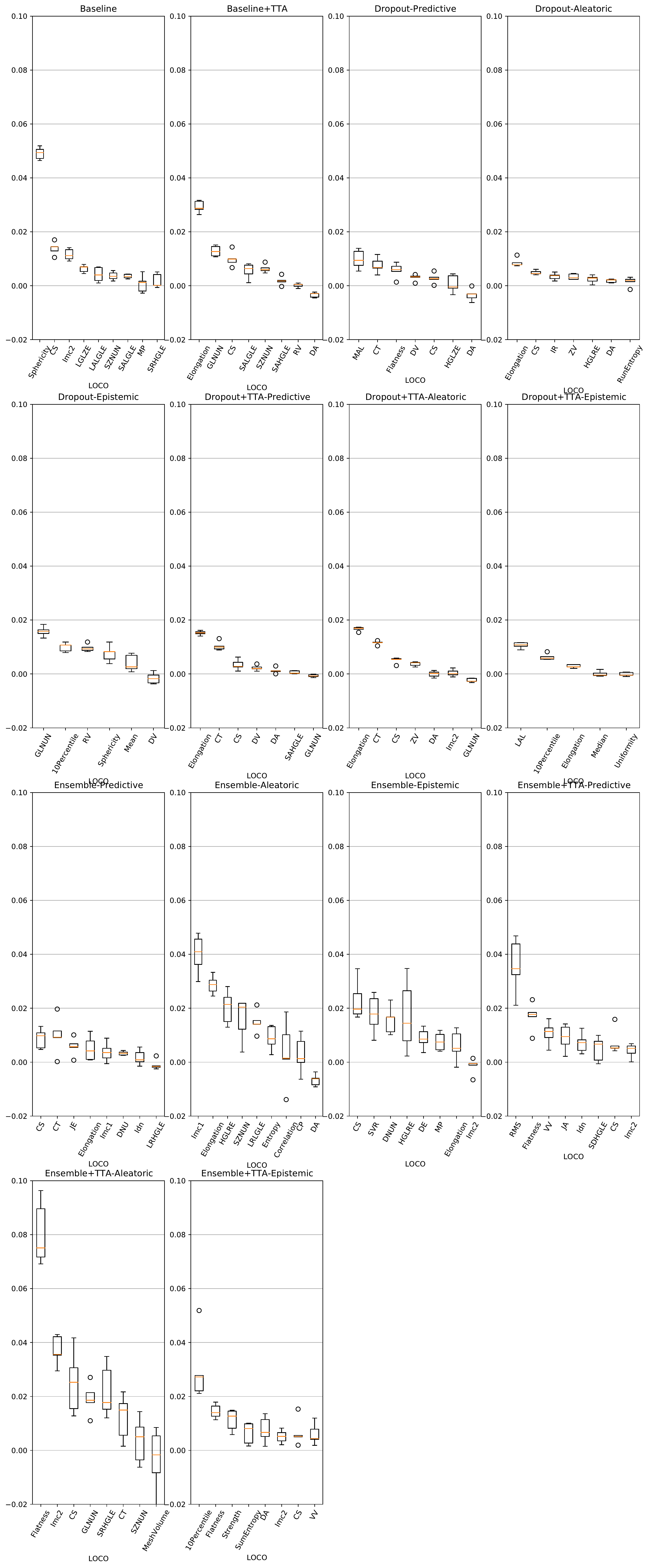}
    \caption{Feature importance scores (LOCO) for UMC classifiers}
    \label{fig:feat_imp_umc_all}
\end{figure}

Feature importance scores from the trained classifiers for LiTS and UMC dataset are shown in Figures \ref{fig:feat_imp_lits_all} and \ref{fig:feat_imp_umc_all}.

\section{Analysis of threshold-based classification}
\label{app:logsum}

\begin{figure}
    \centering
    \includegraphics[width=1.0\linewidth]{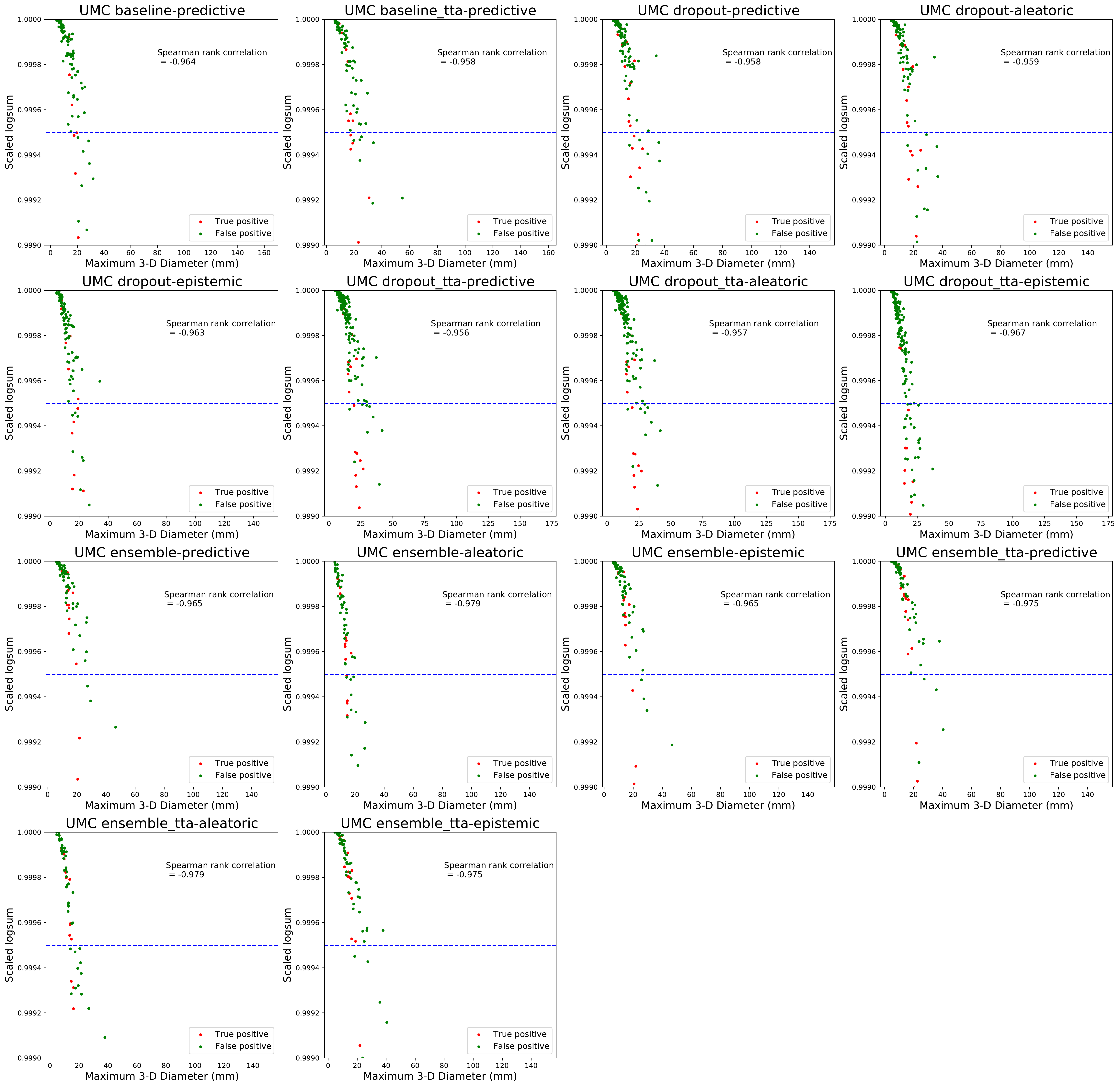}
    \caption{Scatter plot of the (scaled) log-sum aggregate calculated over predicted lesions, and maximum lesion diameter for the UMC test-set. The spearman rank correlation is reported. The detected lesions above the dotted blue line are classified as false positives.}
    \label{fig:logsum_umc}
\end{figure}

\begin{figure}
    \centering
    \includegraphics[width=1.0\linewidth]{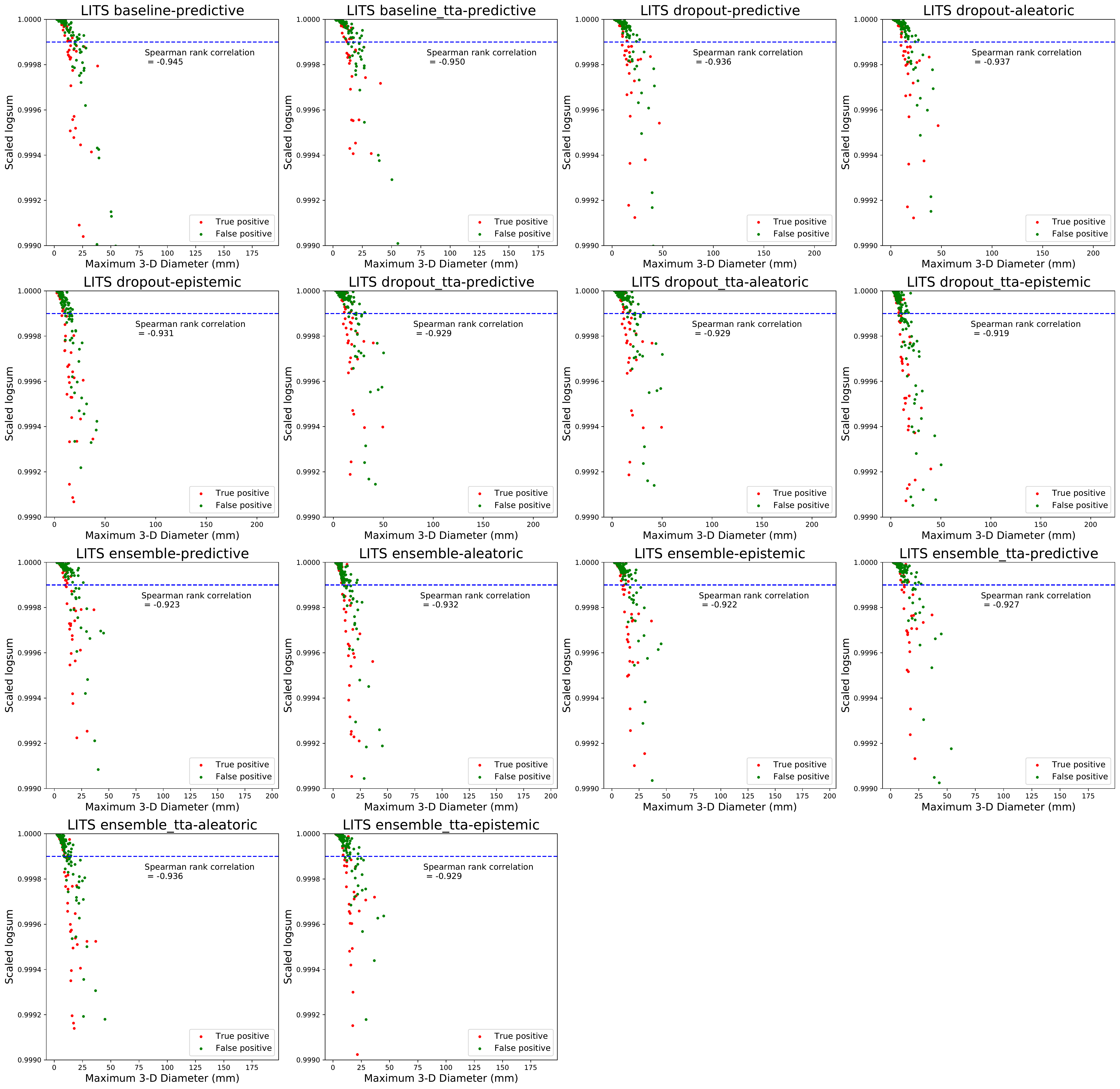}
    \caption{Scatter plot of the (scaled) log-sum aggregate calculated over predicted lesions, and maximum lesion diameter for the LiTS test-set. The spearman rank correlation is reported. The detected lesions above the dotted blue line are classified as false positives.}
    \label{fig:logsum_lits}
\end{figure}

Figures \ref{fig:logsum_umc} and \ref{fig:logsum_lits} show the strong correlation between the log-sum of per-voxel uncertainties and lesion size for both datasets. \cite{nair_exploring_2020} reported that the log-sum computed for smaller lesions is higher, and most false positives are small, therefore log-sum can be used to successfully filter them out. Therefore, the uncertainty log-sum is a proxy for lesion size, as we can see from the high negative correlation between the log-sum and lesion diameter. Therefore, even though it seems that log-sum of the uncertainty over the lesion volume is used to filter false positives, given the high correlation with respect to the size, our observation about the influence of lesion shape on false-positive classification holds.

\end{document}